\documentclass[a4paper,12pt]{article}
\include{sc3conf.sty}
\usepackage{amsmath,amssymb,amscd}
\usepackage{wrapfig,graphicx}

\begin{document}

\begin{center}
{\bf Phenomenological aspects of Supersymmetry: SUSY models
and electroweak symmetry breaking}
\end{center}

\medskip

\begin{center}

{\bf Roman Nevzorov\footnote{On leave of absence from the Theory Department,
ITEP, Moscow, Russia}}

\bigskip

Department of Physics and Astronomy, University of Hawaii, \\
Honolulu, HI 96822, USA
%University of Hawaii

\end{center}

\bigskip
These lectures are a very brief introduction to low energy supersymmetry (SUSY). 
The approach to the construction of SUSY Lagrangians based on the superfield formalism 
is considered. The minimal supersymmetric standard model (MSSM) is specified.
The breakdown of gauge symmetry and Higgs phenomenology within the simplest 
SUSY extensions of the standard model (SM) are briefly reviewed. The upper bound on 
the mass of the lightest Higgs boson and little hierarchy problem in SUSY models 
are discussed.

\section{Introduction}
As is well known, the Standard Model (SM) describes perfectly the major
part of all experimental data measured in earth based experiments.
The Lagrangian of the SM is invariant under Poincare group and
$SU(3)_C\times SU(2)_W\times U(1)_Y$ gauge symmetry transformations.
The Poincare group is an extension of Lorentz group
that includes time and space translations
$$
|\Psi>\to \exp\{-i\hat{H}t\}|\Psi>,\qquad\qquad
|\Psi>\to \exp\{i\hat{\mathbf{P}}\cdot\mathbf{x}\}|\Psi>,
$$
where $\hat{H}$ is a Hamiltonian and $\hat{\mathbf{P}}$ are momentum
operators. 

The transformations of Lorentz group involve rotations 
about three axes and Lorentz boosts along them. Thus there are three generators
of Lorentz group $J_a$ which are associated with the rotations around
three different axes and there are three other generators $K_a$ of this 
group which correspond to the Lorentz boosts along these axes.
Then Lorentz transformations of spin $J$ particle are given by
$$
|J> \to \exp\{i\left(J_a\theta_a+K_b\omega_b\right)\}|J>,
$$
where $\theta_a$ are three rotation angles, while $\omega_b$ are parameters  
which are related to the boost velocity $v$. When $\omega_2=\omega_3=0$ the
parameter $\omega_1$ is defined as $\mbox{tanh}\,\omega_1\,= v$.
The commutation relations between the generators of the Lorentz group 
can be written as
\begin{equation}
\left[J_a, J_b\right]=i\varepsilon_{abc}J_c\,,\qquad 
\left[J_a, K_b\right]=i\varepsilon_{abc}K_c\,,\qquad
\left[K_a, K_b\right]=-i\varepsilon_{abc}J_c\,,
\label{intr2}
\end{equation}
where $a,b,c=1,2,3$. 

It is convenient to introduce two linear combinations
of the generators of Lorentz group $L_a=\frac{1}{2}\left(J_a+iK_a\right)$ 
and $N_a=\frac{1}{2}\left(J_a-iK_a\right)$. $L_a$ and $N_a$ obey two separate 
$SU(2)$ algebras
$$
\left[L_a, L_b\right]=i\varepsilon_{abc}L_c\,,\qquad \left[N_a, N_b\right]=i\varepsilon_{abc}N_c\,,\qquad 
\left[L_a, N_b\right]=0\,. 
$$
As a result the representations of the Lorentz group are classified by two half integer 
numbers $(j_1,\,j_2)$ which are associated with the representations of these two 
$SU(2)$ algebras. The representations $(j,j)$ correspond to integer spin particles
with spin $2j$. The simplest spinor representations are $\psi_L=(1/2,\,0)$ and $\psi_R=(0,\,1/2)$.
The states $\psi_L$ and $\psi_R$ transform differently under the Lorentz transformations. 
Indeed, in the fundamental representation $L_a=\frac{1}{2}\sigma_a$ and $N_b=\frac{1}{2}\sigma'_a$ 
where $\sigma_a$ and $\sigma'_a$ are $2\times 2$ Pauli matrices. Then in the case of
$(1/2,\,0)$ representation $J_a=\frac{1}{2}\sigma_a$ and $K_a=-\frac{i}{2}\sigma_a$ so that 
\begin{equation}
\Psi_L\to\exp\left\{i\frac{\sigma_a}{2}\theta_a+\frac{\sigma_b}{2}\omega_b\right\}\Psi_L\,.
\label{intr3}
\end{equation}
For $(0,\,1/2)$ representation we have $J_a=\frac{1}{2}\sigma'_a$, $K_a=\frac{i}{2}\sigma'_a$
and
\begin{equation} 
\Psi_R\to\exp\left\{i\frac{\sigma'_a}{2}\theta_a-\frac{\sigma'_b}{2}\omega_b\right\}\Psi_R\,.
\label{intr4}
\end{equation}
Thus from Eqs.~(\ref{intr3}) and (\ref{intr4}) it follows that $\psi_L$ and $\psi_R$ transform 
differently under the Lorentz boosts. 

Two spinor representations $(1/2,\,0)$ and $(0,\,1/2)$ are 
called left--handed and right--handed Weyl spinors. The direct sum of the two Weyl spinor states
$(1/2,\,0)\oplus(0,\,1/2)$ describes massive spin $1/2$ (Dirac) state
$$
\Psi(x)=\left(
\begin{array}{c}
\psi_{\alpha}(x)\\[1mm]
\overline{\chi}^{\,\Dot{\alpha}}(x)
\end{array}
\right)\,.
$$
In a Weyl representation $\psi_{\alpha}(x)$ and $\overline{\chi}^{\,\Dot{\alpha}}(x)$
are associated with $(1/2,\,0)$ and $(0,\,1/2)$ spinor representations, i.e. 
left--handed and right--handed Weyl spinors. In the case of massless particle $\psi(x)$ and 
$\overline{\chi}(x)$ have negative and positive helicity. Under parity transformations 
$\psi(x)\leftrightarrow\overline{\chi}(x)$. Charge conjugation results in
$$
\Psi^C=C\,\overline{\Psi}^{\small T}=\left(
\begin{array}{cc}
i\sigma_2 & 0\\[1mm]
0         & -i\sigma_2
\end{array}
\right)
\left(
\begin{array}{c}
\overline{\chi}^{\,\Dot{\beta}\,\dagger}\\[1mm]
\psi^{\dagger}_{\beta}
\end{array}
\right)=\left(
\begin{array}{cc}
\varepsilon_{\alpha\beta} & 0\\[1mm]
0         & \varepsilon^{\Dot{\alpha}\Dot{\beta}}
\end{array}
\right)
\left(
\begin{array}{c}
\overline{\chi}^{\,\Dot{\beta}\,\dagger}\\[1mm]
\psi^{\dagger}_{\beta}
\end{array}
\right)=\left(
\begin{array}{c}
\chi_{\alpha}\\[1mm]
\overline{\psi}^{\,\Dot{\alpha}}
\end{array}
\right),
$$
where $(i\sigma_2 \overline{\chi}^{\dagger})_{\alpha}=\chi_{\alpha}$ and 
$(-i\sigma_2 \psi^{\dagger})^{\,\Dot{\alpha}}=\overline{\psi}^{\,\Dot{\alpha}}$.
The states $\chi_{\alpha}$ and $\overline{\psi}^{\,\Dot{\alpha}}$ transform under 
Lorentz transformations as $(1/2,\,0)$ and $(0,\,1/2)$.
In the Weyl representation the Lagrangian that describes free Dirac fermion 
with mass $m$ can be written as
$$
{\cal L}_D=i\biggl(\chi\sigma^{\mu}\partial_{\mu}\bar{\chi}+
\bar{\psi}\bar{\sigma}^{\mu}\partial_{\mu}\psi\biggr)-m\biggl(\chi\psi+\bar{\psi}\bar{\chi}\biggr)\,,
$$
where $\sigma^{\mu}=(1,\sigma^{i})$,\, $\overline{\sigma}^{\mu}=(1,-\sigma^{i})$
and $\sigma^i$ are Pauli matrices. From now on we use abbreviated notations
$$
\begin{array}{c}
\chi\psi=\chi_{\alpha}\varepsilon^{\alpha\beta}\psi_{\beta}=\chi^{\beta}\psi_{\beta}\,,\qquad\qquad
\bar{\psi}\bar{\chi}=\bar{\psi}^{\Dot{\alpha}}\varepsilon_{\Dot{\alpha}\Dot{\beta}}\bar{\chi}^{\Dot{\beta}}
=\bar{\psi}_{\Dot{\beta}}\bar{\chi}^{\Dot{\beta}}\,,\\[2mm]
\chi\sigma^{\mu}\partial_{\mu}\bar{\chi}=\chi^{\alpha}\sigma^{\mu}_{\alpha\Dot{\alpha}}
\partial_{\mu}\bar{\chi}^{\Dot{\alpha}}\,,\qquad\qquad
\bar{\psi}\bar{\sigma}^{\mu}\partial_{\mu}\psi=\bar{\psi}_{\Dot{\alpha}}(\sigma^{\mu})^{\Dot{\alpha}\alpha}
\partial_{\mu}\psi_{\alpha}\,.
\end{array}
$$
A Majorana spinor is defined as one which is equal to its own charge conjugate, i.e.
$\Psi^T_{M}(x)=\left(\psi_{\alpha}(x),\,\overline{\psi}^{\Dot{\alpha}}(x)\right)$\,.
In the Weyl representation the Lagrangian that describes free Majorana fermion
with mass $m$ takes the form
$$
{\cal L}_M=i(\psi\sigma^{\mu}\partial_{\mu}\bar{\psi})-\frac{m}{2}\biggl(\psi\psi+\bar{\psi}\bar{\psi}\biggr)\,,
$$

The six generators of the Lorentz group $J_a$ and $K_a$ can be presented as 
antisymmetric tensor $M_{\mu\nu}=-M_{\nu\mu}$. The components of $M_{\mu\nu}$
are $K_a=M_{0a}$ and $J_a=\frac{1}{2}\varepsilon_{abc}M_{bc}$.
$M_{\mu\nu}$ is called angular momentum tensor. The translation operators 
$\hat{P}_{\mu}=(\hat{H},\hat{P}_1,\hat{P}_2,\hat{P}_3)$ and the angular
momentum tensor $M_{\mu\nu}$ form a complete set of generators of the
Poincare group which obey algebra
\begin{equation}
\begin{array}{c}
\left[M_{\mu\nu},M_{\rho\sigma}\right]=i\left(g_{\nu\rho}M_{\mu\sigma}-g_{\mu\rho}M_{\nu\sigma}-
g_{\nu\sigma}M_{\mu\rho}+g_{\mu\sigma}M_{\nu\rho}\right)\,,\\[2mm]
\left[\hat{P}_{\mu},\,\hat{P}_{\nu}\right]=0\,,\qquad\qquad
\left[M_{\mu\nu},\hat{P}_{\lambda}\right]=i\left(g_{\nu\lambda}\hat{P}_{\mu}-g_{\mu\lambda}\hat{P}_{\nu}\right)\,.
\end{array}
\label{1}
\end{equation}

The generators $T^a$ associated with $SU(3)_C\times SU(2)_W\times U(1)_Y$
gauge symmetry (internal symmetry group) commute with the generators of
Poincare group. It was established experimentally that $SU(2)_W\times U(1)_Y$
symmetry is broken. The corresponding massive $W^{\pm}$ and $Z$ bosons were
discovered more than 30 years ago. Their properties have been studied in great
detail both theoretically and experimentally. Quarks and gluons that participate
in the strong interactions are confined inside mesons and baryons and therefore
cannot be observed directly. Nevertheless theory of strong interactions based on
$SU(3)_C$ provides a good description for the spectrum of mesons and baryons,
$e^{+}e^{-}$ annihilation data, deep inelastic scattering etc.

Nowadays Higgs boson remains the only missing piece of the SM that has
not been discovered yet. It plays a key role in the SM and its extensions. Higgs 
field acquires vacuum expectation value (VEV) breaking electroweak (EW) symmetry
and generating masses of bosons and fermions.

Although the SM describes the major part of experimental data with high
accuracy it does not allow to incorporate consistently gravitational
interactions. Indeed, in order to achieve the unification of gauge interactions
with gravity we need to combine Poincare and internal symmetries. At the same
time according to the Coleman-Mandula theorem the most general symmetry which
quantum field theory can have is a tensor product of the Poincare group
and an internal group \cite{Coleman:1967ad}. The Coleman-Mandula theorem can
be overcome within graded Lie algebras that have the following structure
$$
\left[\hat{B},\hat{B}\right]=\hat{B},\qquad \left[\hat{B},\hat{F}\right]=\hat{F},\qquad
\left\{\hat{F},\hat{F}\right\}=\hat{B},
$$
where $\hat{B}$ and $\hat{F}$ are bosonic and fermionic generators.
Graded Lie algebras that contain the Poincare algebra are called
supersymmetries. The simplest $N=1$ supersymmetry (SUSY) involves 
a single Weyl spinor operator $Q_{\alpha}$ and its complex conjugate 
$Q_{\alpha}^{\dagger}=\overline{Q}_{\,\Dot{\alpha}}$.
These operators change the spin of the state, i.e.
$$
Q_{\alpha}|fermion>=|boson>,\qquad \overline{Q}_{\dot{\alpha}}|boson>=|fermion>.
$$
The $N=1$ SUSY algebra is Poincare algebra (\ref{1}) plus
\begin{equation}
\begin{array}{ll}
\left\{Q_{\alpha},\overline{Q}_{\dot{\alpha}}\right\}=2\,\sigma^{\mu}_{\alpha\dot{\alpha}}\hat{P}_{\mu},\qquad
&\left\{Q_{\alpha},Q_{\beta}\right\}=\left\{\overline{Q}_{\dot{\alpha}},\overline{Q}_{\dot{\alpha}}\right\}=0,\\[2mm]
\left[\hat{P}_{\mu},Q_{\alpha}\right]=0,\qquad
&\left[M^{\mu\nu},Q_{\alpha}\right]=-i\left(\sigma^{\mu\nu}\right)^{\beta}_{\alpha} Q_{\beta},\\[2mm]
\left[\hat{P}^{\mu},\overline{Q}_{\dot{\alpha}}\right]=0,\qquad
&\left[M^{\mu\nu},\overline{Q}^{\dot{\alpha}}\right]=
-i\left(\sigma^{\mu\nu}\right)^{\dot{\alpha}}_{\dot{\beta}} \overline{Q}^{\dot{\beta}},
\end{array}
\label{2}
\end{equation}
where $\sigma^{\mu\nu}=\frac{1}{4}\left(\sigma^{\mu}\overline{\sigma}^{\nu}
-\sigma^{\nu}\overline{\sigma}^{\mu}\right)$, and
$\overline{\sigma}^{\mu\nu}=\frac{1}{4}\left(\overline{\sigma}^{\mu}\sigma^{\nu}
-\overline{\sigma}^{\nu}\sigma^{\mu}\right)$. The local version of SUSY (supergravity) 
leads to a partial unification of gauge interactions with gravity \cite{susy-grav1}-\cite{susy-grav3}.

This course of lectures is a very basic introduction to $N=1$ supersymmetry. 
In section 2 chiral and vector superfields are introduced. In section 3 these 
superfields are used for the construction of SUSY Lagrangians. The minimal 
supersymmetric standard model (MSSM) is specified in section 4. In section 5
the breakdown of gauge symmetry and Higgs phenomenology within the simplest
SUSY extensions of the SM are considered. 

Nowadays there are many excellent books \cite{b1}--\cite{b6}, introductions \cite{intr1}--\cite{intr15} 
and reviews \cite{rev1}--\cite{rev4} of supersymmetry available while in this paper only a few aspects 
of SUSY are briefly discussed. Students and early--career researchers, who got really 
interested in supersymmetry, should definitely study some of the books, reviews and/or introductions 
mentioned in the list of references of this lecture notes.

\newpage
\section{Superspace and superfields}

An elegant formulation of supersymmetry transformations and
invariants can be achieved in the framework of superspace.
Superspace differs from the ordinary Euclidean (Minkowski)
space by adding of new fermionic coordinates, $\theta_{\alpha}$ and
$\bar{\theta}_{\dot{\alpha}}$, that change in a certain way under
the enlarged group of transformations. These new coordinates are
anticommuting Grassmann variables transforming as two--component
Weyl spinors:
$$
\{ \theta_{\alpha}, \theta_{\beta} \} = 0\,,\qquad
\{ \bar{\theta}_{\dot \alpha}, \bar{\theta}_{\dot \beta} \} = 0\,,\qquad
\theta_{\alpha}^2 = 0\,,\qquad \bar{\theta}_{\dot \alpha}^2=0\,,
$$
where $\alpha,\,\beta,\,\dot{\alpha},\,\dot{\beta} =1,2$.
Fermionic coordinates arise in a SUSY transformation that
can be constructed in superspace in the same way as an ordinary
translation in the usual space
\begin{equation}
G(x, \theta ,\bar{\theta}) =
e^{\displaystyle i(-x^{\mu}P_{\mu} + \theta Q + \bar{\theta} \bar{Q})}.
\label{ss1}
\end{equation}
Using the Hausdorff formula
$$
e^{A} e^{B}=\exp(A+B+\frac{1}{2}[A,B]+...)\,,
$$
that terminates at the first commutator for the group elements considered here,
one can combine two SUSY transformations
\begin{equation}
G(x^{\mu}, \theta ,\bar{\theta}) G(a^{\mu}, \xi ,\bar{\xi})=
G(x^{\mu}+a^{\mu}-i\xi\sigma^{\mu}\bar{\theta}+i\theta\sigma^{\mu}\bar{\xi} ,
\theta+\xi ,\bar{\theta}+\bar{\xi})\,.
\label{ss2}
\end{equation}
This leads to a supertranslation in superspace
\begin{equation}
\begin{array}{ccl}
x_{\mu} & \rightarrow & x_{\mu} + a_{\mu}+ i\theta\sigma_{\mu}\bar{\xi}
-i\xi\sigma_{\mu}\bar{\theta}\,, \\[2mm]
\theta & \rightarrow & \theta + \xi\,,\\[2mm]
\bar{\theta} & \rightarrow & \bar{\theta} + \bar{\xi}\,,
\end{array}
\label{ss3}
\end{equation}
where $\xi $ and $\bar{\xi}$ play a role of Grassmannian transformation parameters.
Taking these parameters to be local, or space-time dependent, one gets a local
translations that result in a theory of (super) gravity.

Whereas an ordinary field is a function of the space--time coordinates $x$ only,
a superfield $S(x, \theta ,\bar{\theta})$ is also a function of anticommuting
Grassmann variables $\theta_{\alpha}$ and $\bar{\theta}_{\dot{\alpha}}$. The fields
of the supermultiplet then arise as the coefficients in an expansion of
$S(x, \theta ,\bar{\theta})$ in powers of $\theta_{\alpha}$ and
$\bar{\theta}_{\dot{\alpha}}$.
Assuming that $a_{\mu}$, $\xi $ and $\bar{\xi}$ are infinitesimally small
we can Taylor expand $S(x'^{\mu}, \theta' ,\bar{\theta}')$
\begin{equation}
\begin{array}{c}
S(x^{\mu}+a^{\mu}-i\xi\sigma^{\mu}\bar{\theta}+i\theta\sigma^{\mu}\bar{\xi} ,
\theta+\xi ,\bar{\theta}+\bar{\xi})\simeq S(x^{\mu}, \theta ,\bar{\theta})+\\[2mm]
\displaystyle (a^{\mu}-i\xi\sigma^{\mu}\bar{\theta}+i\theta\sigma^{\mu}\bar{\xi})
\frac{\partial S}{\partial x^{\mu}}+
\xi^{\alpha} \frac{\partial S}{\partial \theta^{\alpha}}+
\bar{\xi}_{\dot{\alpha}}\frac{\partial}{\bar{\theta}_{\dot{\alpha}}}+...
\end{array}
\label{ss4}
\end{equation}
By comparing the coefficients of the infinitesimal parameters $a_{\mu}$, $\xi $
and $\bar{\xi}$ one can see that the action of the SUSY algebra on superfields
\begin{equation}
S(x^{\mu}, \theta ,\bar{\theta}) \rightarrow
e^{\displaystyle i(-a^{\mu}P_{\mu} + \xi Q + \bar{\xi} \bar{Q})}
S(x^{\mu}, \theta ,\bar{\theta})
\label{ss5}
\end{equation}
is generated by
\begin{equation}
P_{\mu}=i\partial_{\mu},\qquad
iQ_{\alpha} =\frac{\partial }{\partial \theta^{\alpha}
}-i\sigma^{\mu}_{\alpha \dot{\alpha}}\bar{\theta}^{\dot{\alpha}}\partial_{\mu}\,,\qquad
i\bar{Q}_{\dot \alpha} =-\frac{\partial}{\partial \bar{\theta}^{\dot{\alpha}}
}+i\theta^{\alpha}\sigma^\mu_{\alpha \dot{\alpha}}\partial_\mu .
\label{ss6}
\end{equation}

The general superfield $S(x^{\mu}, \theta ,\bar{\theta})$ can be expanded as
a power series in $\theta_{\alpha}$ and $\bar{\theta}_{\dot{\alpha}}$
involving not more than two powers of $\theta_{\alpha}$ and
$\bar{\theta}_{\dot{\alpha}}$ since $\theta_{\alpha}$ and
$\bar{\theta}_{\dot{\alpha}}$ are two--component Grassmann variables.
Scalar superfield $S(x^{\mu}, \theta ,\bar{\theta})$ is the simplest
representation of the Super-Poincare group. However such superfield
is a reducible representation of SUSY. Irreducible representations
of the SUSY algebra are obtained by imposing constraints which are
covariant under the supersymmetry algebra.

In this context it is convenient to define fermionic derivatives
\begin{equation}
D_{\alpha} = \displaystyle\frac{\partial}{\partial \theta^{\alpha}} +
i \sigma^{\mu}_{\alpha\dot{\alpha}} \bar{\theta}^{\dot{\alpha}}\partial_{\mu}\,,\qquad
\bar{D}_{\dot{\alpha}} =\displaystyle -\frac{\partial}{
\partial \bar{ \theta}^{\dot{\alpha}}} -
i \theta^{\alpha} \sigma^{\mu}_{\alpha\dot{\alpha}}\partial_{\mu}\,.
\label{ss7}
\end{equation}
The derivatives $D_{\alpha}$ and $\bar{D}_{\dot{\alpha}}$ anticommute with
the generators of the SUSY algebra, i.e.
\begin{equation}
\{ D_{\alpha} , Q_{\beta} \} = \{ D_{\alpha} , \bar{Q}_{\dot{\beta}} \} =
\{ \bar{D}_{\dot{\alpha}} , Q_{\beta} \} = \{ \bar{D}_{\dot{\alpha}} , \bar{Q}_{\dot{\beta}} \}=0\,.
\label{ss8}
\end{equation}
On the other hand they obey algebra
\begin{equation}
\{ D_{\alpha} , \bar{D}_{\dot{\alpha}} \} = 2 i \sigma^{\mu}_{\alpha\dot{\alpha}}\partial_{\mu}\,,\qquad
\{ D_{\alpha} , D_{\beta} \} = \{ \bar{D}_{\dot{\alpha}} , \bar{D}_{\dot{\beta}} \} =0\,.
\label{ss9}
\end{equation}
The superspace covariant derivatives $D_{\alpha}$ and $\bar{D}_{\dot{\alpha}}$ can be used to impose
covariant constraints on superfields because they commute with $\xi Q +\bar{\xi} \bar{Q}$
which occurs in SUSY transformations (\ref{ss5}). Indeed, if superfield
$\Phi(x^{\mu}, \theta, \bar{\theta})$ satisfies the covariant condition
\begin{equation}
\bar{D}_{\dot{\alpha}} \Phi(x^{\mu}, \theta, \bar{\theta})=0\,,
\label{ss10}
\end{equation}
then after SUSY transformation the corresponding superfield still satisfies
constraint (\ref{ss10}). Superfield $\Phi(x^{\mu}, \theta, \bar{\theta})$ which obey
condition (\ref{ss10}) is called a chiral superfield.

It is worth to point out that
\begin{equation}
\bar{D}_{\dot{\alpha}} \theta_{\alpha} =0\,,\qquad \bar{D}_{\dot{\alpha}} y^{\mu}=0\,,
\label{ss11}
\end{equation}
where $y^{\mu}=x^{\mu}+i\theta\sigma^{\mu}\bar{\theta}$. Thus any function of $\theta$ and
$y^{\mu}$ satisfies the covariant condition (\ref{ss10}). Expanding $\Phi(y^{\mu},\theta)$
in powers of the two-component Grassmann variable $\theta$ gives
\begin{equation}
\Phi (y, \theta ) =  \phi(y) + \sqrt{2}\, \theta \psi (y) + \theta \theta F(y)\,,
\label{ss12}
\end{equation}
where $\phi$ is a complex scalar field, $F$ is an auxiliary complex scalar field and
$\psi$ is a left--handed Weyl spinor field. The coefficients in expansion (\ref{ss12})
are called the components of a superfield. The general expansion of chiral superfield
(\ref{ss12}) in component fields around $x^{\mu}$ is
\begin{equation}
\begin{array}{c}
\displaystyle \Phi(x^{\mu}, \theta, \bar{\theta}) = \phi(x) + \sqrt{2}\,\theta\psi (x) + (\theta\theta) F(x)
+i(\theta \sigma^{\mu} \bar{\theta}) \partial_{\mu}\phi(x)\\[2mm]
\displaystyle - \frac{i}{\sqrt{2}} (\theta\theta) (\partial_{\mu} \psi (x) \sigma^{\mu} \bar{\theta})
- \frac{1}{4} (\theta \theta) (\bar{\theta} \bar{\theta}) \partial_{\mu} \partial^{\mu} \phi(x)\,.
\end{array}
\label{ss13}
\end{equation}
The fields $\phi(x)$ and $\psi(x)$ are called superpartners. From expansion (\ref{ss12})
it follows that chiral superfield has the same number of bosonic and fermionic degrees
of freedom. The mass dimensions of scalars and spinors in Eq.~(\ref{ss12}) are
$[\phi]=1$, $[\psi]=3/2$, $[F]=2$ and $[\theta]=-1/2$ whereas chiral superfield has
mass dimension 1, i.e. $[\Phi]=1$

One can also construct an antichiral (conjugate) superfield
$\Phi^{\dagger}(x^{\mu}, \theta, \bar{\theta})$ that has the component field
expansion
\begin{equation}
\begin{array}{c}
\displaystyle \Phi^{\dagger}(x^{\mu}, \theta, \bar{\theta}) = \phi^{\dagger}(x) +
\sqrt{2}\,\bar{\theta}\bar{\psi} (x) + (\bar{\theta}\bar{\theta}) F^{\dagger}(x)
-i(\theta \sigma^{\mu} \bar{\theta}) \partial_{\mu}\phi^{\dagger}(x)\\[2mm]
\displaystyle + \frac{i}{\sqrt{2}} (\bar{\theta}\bar{\theta}) (\theta\sigma^{\mu} \partial_{\mu} \bar{\psi} (x))
- \frac{1}{4} (\theta \theta) (\bar{\theta} \bar{\theta}) \partial_{\mu} \partial^{\mu} \phi^{\dagger}(x)\,.
\end{array}
\label{ss14}
\end{equation}
This superfield obey equation
\begin{equation}
D_{\alpha} \Phi^{\dagger}(x^{\mu}, \theta, \bar{\theta})=0\,.
\label{ss15}
\end{equation}
Since $\Phi(x^{\mu}, \theta, \bar{\theta})$ involves left--handed Weyl spinor $\psi$
while $\Phi^{\dagger}(x^{\mu}, \theta, \bar{\theta})$ contains right--handed  Weyl spinor $\bar{\psi}$,
$\Phi(x^{\mu}, \theta, \bar{\theta})$ and $\Phi^{\dagger}(x^{\mu}, \theta, \bar{\theta})$ are
sometimes called left--handed and right--handed chiral superfields.

The SUSY transformation of chiral superfield (\ref{ss5}) induces transformations
of its components. Under an infinitesimal SUSY transformation the left--handed superfield
changes as follows
\begin{equation}
\begin{array}{l}
\Phi \rightarrow \Phi + \delta \Phi\,,\\[2mm]
\delta\Phi= i(\xi Q + \bar{\xi} \bar{Q})\Phi=
\delta \phi(x) + \sqrt{2}\, \theta \delta\psi (x) + (\theta \theta) \delta F(x) +...
\end{array}
\label{ss16}
\end{equation}
Using the explicit expressions for $Q$ and $\bar{Q}$ (see Eq.~(\ref{ss6})),
we find
\begin{eqnarray}
\delta \phi &=& \sqrt{2}\, \xi \psi\,, \nonumber \\
\delta \psi &=& i \sqrt{2}\, \sigma^\mu \bar{\xi} \partial_{\mu} \phi + \sqrt{2}\, \xi F\,, \\
\delta F &=& i \sqrt{2}\,  \partial_{\mu}\psi \sigma^{\mu} \bar{\xi}\,. \nonumber
\label{ss17}
\end{eqnarray}
As expected the change in the bosonic (fermionic) component of the superfield is proportional
to the fermionic (bosonic) fields. It is important to notice that the change in $F$
component under a SUSY transformation is a total derivative. Therefore this
component of chiral superfield can be used for the construction of SUSY Lagrangians

For the construction of SUSY Lagrangians it is also necessary to study the products of chiral
superfields and $\Phi^{\dagger}_i \Phi_{j}$. Because $Q_{\alpha}$, $\bar{Q}_{\dot{\alpha}}$
and covariant derivatives $D_{\alpha}$ and $\bar{D}_{\dot{\alpha}}$ are linear differential
operators on superspace any product of chiral (antichiral) superfields $\Phi_i\Phi_j$,
$\Phi_i\Phi_j\Phi_k$ etc is again a chiral (antichiral) superfield so that
the covariant condition (\ref{ss10}) is fulfilled. Direct calculation gives
\begin{equation}
\begin{array}{c}
\Phi_i (y, \theta ) \Phi_j (y, \theta ) =  \phi_i(y) \phi_j(y) +
\sqrt{2}\, \theta (\psi_i (y) \phi_j(y) + \phi_i(y) \psi_j(y)) \\[2mm]
+ (\theta \theta) (\phi_i (y) F_j(y) + \phi_j (y) F_i(y) - \psi_i (y) \psi_j(y))\,,
\end{array}
\label{ss18}
\end{equation}
\begin{equation}
\begin{array}{c}
\Phi_i (y, \theta ) \Phi_j (y, \theta ) \Phi_k (y, \theta)=
 \phi_i(y) \phi_j(y) \phi_k(y) +  \sqrt{2}\, \theta (\psi_i (y) \phi_j(y) \phi_k(y) + \\[2mm]
\phi_i(y) \psi_j(y) \phi_k(y) + \phi_i(y) \phi_j(y) \psi_k(y))+
(\theta \theta) (\phi_i(y) \phi_j(y) F_k(y) + \\[2mm]
\phi_i (y) F_j(y) \phi_k(y) + F_i(y) \phi_j(y) \phi_k(y) - \psi_i (y) \psi_j(y) \phi_k(y)\\[2mm]
- \psi_i (y) \psi_k(y) \phi_j(y) - \psi_j (y) \psi_k(y) \phi_i(y))\,,
\end{array}
\label{ss19}
\end{equation}
For any arbitrary function of chiral superfields one gets
\begin{eqnarray}
W(\Phi_k)&=& W(\phi_k+\sqrt{2}\theta \psi_k+\theta\theta F_k)
=W(\phi_k)+\frac{\partial W(\phi_k)}{\partial \phi_i}\sqrt{2}\theta \psi_i \nonumber \\
&+&\theta \theta
\left(\frac{\partial W(\phi_k)}{\partial \phi_i}F_i -
\frac{1}{2}\frac{\partial^2 W(\phi_k)}{\partial \phi_i\partial
\phi_j}\psi_i\psi_j \right).
\label{ss20}
\end{eqnarray}

On the other hand the product of the left--handed and right--handed chiral superfields
is not a chiral superfield since
\begin{equation}
\begin{array}{rcl}
\tilde{V}&=&\Phi^{\dagger}_i(y,\theta)\Phi_j(y,\theta)=\phi^{\dagger}_i(y) \phi_j(y)+
\sqrt{2}\, (\theta \psi_j(y)) \phi^{\dagger}_i(y) + \\[2mm]
&&\sqrt{2}\, (\bar{\theta} \bar{\psi}_i(y)) \phi_j(y)
+2 (\bar{\theta} \bar{\psi}_i(y))(\theta \psi_j(y))+ F_j(y)\phi^{\dagger}_i(y)(\theta\theta)+\\[2mm]
&&F^{\dagger}_i(y)\phi_j(y)(\bar{\theta}\bar{\theta})+
\sqrt{2}\,(\theta\theta)(\bar{\theta} \bar{\psi}_i(y)) F_j(y)+\\[2mm]
&&\sqrt{2}\,(\bar{\theta}\bar{\theta})(\theta \psi_j(y)) F^{\dagger}_i(y)+
(\bar{\theta}\bar{\theta})(\theta\theta) F^{\dagger}_i(y) F_j(y)
\end{array}
\label{ss21}
\end{equation}
$\tilde{V}$ can not be reduced to the form (\ref{ss12}).
At the same time the superfield $\tilde{V}(x^{\mu}, \theta, \bar{\theta})$
satisfies the constraint
\begin{equation}
\tilde{V}(x^{\mu}, \theta, \bar{\theta})=\tilde{V}^{\dagger}(x^{\mu}, \theta, \bar{\theta})\,,
\label{ss22}
\end{equation}
which is preserved under SUSY transformations. The superfields that obey
condition (\ref{ss22}) are called vector superfields. Note that
$(\bar{\theta}\bar{\theta})(\theta\theta)$ component of $\tilde{V}(x^{\mu}, \theta, \bar{\theta})$
is proportional to $F^{\dagger}_i F_j$. This is an indication that the coefficient of the
$(\bar{\theta}\bar{\theta})(\theta\theta)$ term is a new auxiliary field.
This auxiliary field is called $D$. One can anticipate that the new auxiliary field
can be also useful in constructing SUSY Lagrangians. After substituting for $y$ we find
that
\begin{equation}
\begin{array}{c}
\displaystyle \left[\Phi^{\dagger}_i\Phi_j\right]_D = F^{\dagger}_i F_j+
\frac{1}{2}\partial_{\mu}\phi^{\dagger}_i\partial^{\mu}\phi_j -
\frac{1}{4}\phi_i^{\dagger}\partial_{\mu}\partial^{\mu}\phi_j -
\frac{1}{4} (\partial_{\mu}\partial^{\mu} \phi_i^{\dagger}) \phi_j\\[2mm]
\displaystyle +\frac{i}{2} \bar{\psi}_i \bar{\sigma}^{\mu}\partial_{\mu}\psi_j-
\frac{i}{2} (\partial_{\mu}\bar{\psi}_i)\bar{\sigma}^{\mu}\psi_{j}\,.
\end{array}
\label{ss23}
\end{equation}

Another example of vector superfield that can be constructed using
left--handed and right--handed chiral superfields is
\begin{equation}
\begin{array}{rcl}
V'(x, \theta, \bar{\theta})&=&i(\Phi-\Phi^{\dagger})=i(\phi(x)-\phi^{\dagger}(x))+
i\sqrt{2}(\theta \psi(x) - \bar{\theta} \bar{\psi}(x))\\[2mm]
&&\displaystyle +i(\theta\theta F(x)-\bar{\theta}\bar{\theta} F^{\dagger}(x)) -
(\theta\sigma^{\mu}\bar{\theta})\partial_{\mu}(\phi(x)+\phi^{\dagger}(x))\\[2mm]
&&\displaystyle -\frac{1}{\sqrt{2}}(\theta\theta)(\bar{\theta}\bar{\sigma}^{\mu}\partial_{\mu}\psi(x))
+\frac{1}{\sqrt{2}}(\bar{\theta}\bar{\theta})(\theta\sigma^{\mu}\partial_{\mu}\bar{\psi}(x))\\[3mm]
&&\displaystyle -\frac{i}{4}(\theta\theta)(\bar{\theta}\bar{\theta})
\partial_{\mu}\partial^{\mu}(\phi(x)-\phi^{\dagger}(x))\,.
\end{array}
\label{ss24}
\end{equation}
We will use the explicit expression for vector superfield $V'(x, \theta, \bar{\theta})$
when we consider (super) gauge transformations.

Without loss of generality any Lorentz invariant superfield may be written
in the form
\begin{equation}
\begin{array}{c}
F(x, \theta, \bar{\theta})=f(x)+\theta\chi'(x)+\bar{\theta}\bar{\chi}+\theta\theta m(x)+
\bar{\theta}\bar{\theta} \tilde{m}(x)+(\theta\sigma^{\mu}\bar{\theta})V_{\mu}(x)\\[2mm]
+\theta\theta\bar{\theta}\bar{\lambda}(x)+\bar{\theta}\bar{\theta}\theta\lambda'(x)+
\theta\theta\bar{\theta}\bar{\theta}d(x)\,,
\end{array}
\label{ss25}
\end{equation}
where $f(x),\, m(x),\,\tilde{m}(x),\, d(x)$ are scalar fields, $V_{\mu}(x)$ is a vector
field, and $\chi'(x),\,\chi(x),\,\lambda(x),\,\lambda'(x)$ are Weyl spinor fields.
Because all these fields can be complex the superfield $F(x, \theta, \bar{\theta})$
involves eight complex bosonic and fermionic degrees of freedom.
If we require that $F(x, \theta, \bar{\theta})$ is real, i.e. it satisfies condition (\ref{ss22}),
then
\begin{equation}
\begin{array}{c}
f(x)=f^{\dagger}(x)\,,\quad  m(x)=\tilde{m}^{\dagger}(x)\,,\quad d(x)=d^{\dagger}(x)\,,\quad
V_{\mu}(x)=V^{\dagger}_{\mu}(x)\,,\\[2mm]
\chi'(x)=\chi(x)\,,\quad \lambda(x)=\lambda'(x)\,.
\end{array}
\label{ss26}
\end{equation}
Thus, through the constraint (\ref{ss22}) the eight complex bosonic and fermionic
degrees of freedom in Eq.~(\ref{ss25}) are reduced to eight real bosonic and fermionic
degrees of freedom. From Eqs.~(\ref{ss25})--(\ref{ss26}) it becomes clear that vector
superfield contains a real vector field which can play a role of gauge vector boson.
Therefore this superfield has to be used to formulate a supersymmetric version of
quantum electrodynamics (QED), for example, which involves photon.

It is convenient to rewrite vector superfield using special field combinations for the
coefficients of the $\theta\theta\bar{\theta}$, $\bar{\theta}\bar{\theta}\theta$ and
$\theta\theta\bar{\theta}\bar{\theta}$ components, i.e.
\begin{eqnarray}
V(x, \theta, \bar \theta) & = & C(x) + i\theta \chi (x)
-i\bar{\theta}\bar{\chi}(x)  + \frac{i}{2} \theta \theta [M(x) + iN(x)] \nonumber\\
&&-\frac{i}{2} \bar{\theta}\bar{\theta}[M(x)-iN(x)] -
\theta\sigma^{\mu}\bar{\theta}V_{\mu}(x)\\
&&+i\theta\theta\bar{\theta}\biggr[\bar{\lambda}(x)
+\frac{i}{2}\bar{\sigma}^{\mu}\partial_{\mu}\chi (x)\biggl]
- i\bar{\theta}\bar{\theta}\theta \biggr[\lambda(x) +
\frac{i}{2}\sigma^{\mu}\partial_{\mu}\bar{\chi}(x)\biggl]
\nonumber\\
&&+\frac{1}{2} \theta \theta \bar{\theta}\bar{\theta}\biggr[D(x) -
\frac{1}{2}\partial_{\mu}\partial^{\mu} C(x)\biggl]
\nonumber\,,
\label{ss27}
\end{eqnarray}
where $C(x)$, $M(x)$, $N(x)$, $D(x)$ are real scalar fields.
As before, $\chi$ and $\lambda$ are Weyl spinor fields and
$V_{\mu}$ is a real vector field. The dimensions of all these
fields are fixed by requiring that the vector field $V_{\mu}$
has its canonical mass dimension $[V]=1$. Then
\begin{eqnarray}
[C]=0\,,\quad [\chi]=\frac{1}{2}\,,\quad [M]=[N]=1\,,\quad
[\lambda]=\frac{3}{2}\,,\quad [D]=2\,,
\label{ss28}
\end{eqnarray}
while the vector superfield is dimensionless.

In the QED Lagrangian does not change when
$V_{\mu}\rightarrow V_{\mu}-\partial_{\mu}\omega$.
In SUSY QED $V_{\mu}$ and $\omega$ have to be the appropriate
components of SUSY multiplets. Therefore gauge transformation
should be defined in terms of superfields. One can expect that
$\omega$ might appear as a component of the linear superposition
of the left--handed and right--handed chiral superfields
that transform as a vector superfield with respect to
SUSY transformations. One combination of
$\Phi(x,\theta,\bar{\theta})$ and
$\Phi^{\dagger}(x,\theta,\bar{\theta})$ that satisfies this
requirement was mentioned before (see Eq.~(\ref{ss24})). Indeed,
if under $U(1)$ gauge transformation $V(x,\theta,\bar{\theta})$
transforms as
\begin{eqnarray}
V(x,\theta,\bar{\theta}) \rightarrow V(x,\theta,\bar{\theta}) +
i\biggl[\Phi(x,\theta,\bar{\theta}) - \Phi^{\dagger}(x,\theta,\bar{\theta})\biggr]\,,
\label{ss29}
\end{eqnarray}
then the components of vector superfield changes as follows
\begin{eqnarray}
C(x) & \rightarrow & C(x) + i(\phi(x) - \phi^{\dagger}(x))\,,\nonumber \\
\chi(x) & \rightarrow & \chi(x) + \sqrt{2} \psi(x)\,, \nonumber \\
M(x) + iN(x)  & \rightarrow & M(x) + iN(x) + 2F(x)\,, \nonumber \\
V_{\mu}(x) & \rightarrow & V_{\mu}(x) -\partial_{\mu} (\phi(x) + \phi^{\dagger}(x))\,, \label{ss30}\\
\lambda(x) & \rightarrow & \lambda(x)\,, \qquad D(x) \rightarrow  D(x)\,.\nonumber
\end{eqnarray}
From Eq.~(\ref{ss30}) it is easy to see that local $U(1)$ gauge field
transforms as in the QED. Thus Eq.~(\ref{ss29}) can be considered
as a supersymmetric generalization of gauge transformation in SUSY QED.
One can also see that $C(x)$, $M(x)$, $N(x)$ and $\chi(x)$ are not physical
degrees of freedom, since they can be gauged away by a suitable choice of
$(\phi(x) - \phi^{\dagger}(x))$, $\psi(x)$ and $F(x)$ while still leaving
$\omega(x)=(\phi(x) + \phi^{\dagger}(x))$ arbitrary. In this Wess-Zumino
gauge \cite{Wess:1974jb} vector superfield takes the form
\begin{eqnarray}
V(x,\theta,\bar{\theta})=-\theta \sigma^{\mu}\bar{\theta} V_{\mu}(x) + i\theta\theta\bar{\theta}\bar{\lambda}(x)
-i\bar{\theta}\bar{\theta}\theta\lambda (x) + \frac{1}{2}\theta\theta\bar{\theta}\bar{\theta}D(x)\,.
\label{ss31}
\end{eqnarray}
From (\ref{ss30}) it becomes clear that
the fields $\lambda(x)$ and $D(x)$ are gauge invariant whereas
$V_{\mu}(x)$ transforms as in the usual QED. So $V_{\mu}(x)$ should be associated with the gauge
field (photon) while $\lambda(x)$ is called gaugino (photino in SUSY QED). All powers
$V^n(x,\theta,\bar{\theta})$ with $n>2$ vanish in the Wess-Zumino gauge, since they will
involve at least $\theta^3$. The only non--zero power is
$$
V^2(x,\theta,\bar{\theta}) = \frac{1}{2} \theta \theta
\bar{\theta}\bar{\theta} V_{\mu}(x)V^{\mu}(x)\,.
$$

As we have done for the chiral superfield we can now determine the transformation
properties of the component fields of $V(x,\theta,\bar{\theta})$.
Under an infinitesimal SUSY transformation
the vector superfield changes as follows
\begin{equation}
V \rightarrow V + \delta V\,,\qquad \delta V= i(\xi Q + \bar{\xi} \bar{Q})V.
\label{ss32}
\end{equation}
With $Q$ and $\bar{Q}$ given by Eq.~(\ref{ss6}) we find
\begin{eqnarray}
\delta \lambda_{\alpha} &=& -i D \xi_{\alpha}-
\frac{1}{2}(\sigma^{\mu}\bar{\sigma}^{\nu})^{\beta}_{\alpha}\xi_{\beta} V_{\mu\nu}\,,\nonumber \\
\delta V^{\mu} &=& i (\xi\sigma^{\mu}\bar{\lambda}-\lambda\sigma^{\mu}\bar{\xi})-
\partial^{\mu}(\xi\chi+\bar{\xi}\bar{\chi})\,, \label{ss33}\\
\delta D &=& \partial_{\mu}(-\xi\sigma^{\mu}\bar{\lambda}+\lambda\sigma^{\mu}\bar{\xi})\,, \nonumber
\end{eqnarray}
where $V_{\mu\nu}=\partial_{\mu} V_{\nu}-\partial_{\nu} V_{\mu}$. Eqs.~(\ref{ss33})
lead to the following transformation property of the $U(1)$ field strength $V_{\mu\nu}$
\begin{eqnarray}
\delta V^{\mu\nu}=i\partial^{\mu}(\xi\sigma^{\nu}\bar{\lambda}-\lambda\sigma^{\nu}\bar{\xi})
-i\partial^{\nu}(\xi\sigma^{\mu}\bar{\lambda}-\lambda\sigma^{\mu}\bar{\xi}),
\label{ss34}
\end{eqnarray}
that does not depend on the Wel spinor $\chi$. Eqs.~(\ref{ss33})--(\ref{ss34}) imply that
fields $V_{\mu\nu}$, $\lambda$ and $D$ form an irreducible representation of the SUSY
algebra by themselves. Note that the variation of the $D$--field is a total divergence
as in the case of $F$--field. Since total divergences vanish when integrated over the
space time the $F$--components of chiral superfields and the $D$--components of vector
superfields can be used to construct SUSY Lagrangians.

\section{SUSY Lagrangians}

The discussion at the end of the previous section suggests an immediate way to construct
Lagrangians which are invariant under SUSY transformations. In the superfield notation
SUSY invariant Lagrangians are the polynomials of superfields. Then the key observation
for the construction of SUSY theories is that the $F$--term of a chiral superfield
(i.e. the $\theta\theta$ component of a left--handed chiral superfield and the
$\bar{\theta}\bar{\theta}$ component of a right--handed chiral superfield) and
the $D$--term of a vector superfield (i.e. the $\theta\theta\bar{\theta}\bar{\theta}$
component) transform into themselves plus a total derivative under SUSY transformations.
Since the action does not change when Lagrangian ${\cal L}$ changes by a total derivative,
the SUSY invariant Lagrangian can be written as
\begin{eqnarray}
{\cal L} = {\cal L}_D + {\cal L}_F\,,
\label{lag1}
\end{eqnarray}
where ${\cal L}_F$ is made up of $F$--terms and ${\cal L}_D$ is made up of $D$--terms.

\subsection{Lagrangians for chiral superfields}
Let us start with the Lagrangian which contains only a set of chiral superfields and
has no vector supermultiplets (Wess--Zumino model). Then the most general renormalizable
SUSY invariant Lagrangian has the form
\begin{eqnarray}
{\cal L}_{WZ} = {\cal L}_D + {\cal L}_F = \sum_i \biggl[\Phi_i^{\dagger} \Phi_i \biggr]_{D}
+ \biggl([W(\Phi_k)]_{F} + h.c.\biggr) \,,
\label{lag2}
\end{eqnarray}
where $W(\Phi_k)$, which is referred to as the superpotential, must involve only
up to the third power of the superfields $\Phi_k$ to obtain a renormalizable Lagrangian,
i.e.
\begin{eqnarray}
W(\Phi_k)= a_i \Phi_i + \frac{1}{2}\mu_{ij}\Phi_i \Phi_j + \frac{1}{3} y_{ijk} \Phi_i \Phi_j \Phi_k\,.
\label{lag3}
\end{eqnarray}
In Eq.~(\ref{lag3}) the sum over all possible combinations of chiral superfields is
understood while $a_i$, $\mu_{ij}$ and $y_{ijk}$ are constants. One might think that
we could add more terms with products of more than three chiral superfields in the
superpotential. In this context it is worth to note that $[\Phi_i]_F$, $[\Phi_i \Phi_j]_F$
and $[\Phi_i \Phi_j \Phi_k]_F$ have mass dimensions 2, 3 and 4 respectively. As a
consequence the mass dimensions of various couplings in Eq.~(\ref{lag3}) are
$[a_i]=2$, $[\mu_{ij}]=1$ and $[y_{ijk}]=0$ to ensure $[{\cal L}_F]=4$. It is obvious
that $F$--terms involving more factors of $\Phi_i$ will have mass dimension greater
than four resulting in couplings with negative mass dimension. The presence of such
couplings makes SUSY theory non--renormalizable.

The superpotential $W(\Phi_k)$ must contain the products of chiral superfields only.
In other words $W(\Phi_k)$ has to be a holomorphic (analytic) function of $\Phi_k$.
The presence of terms like $\Phi_i\Phi^{\dagger}_j$, $\Phi_i\Phi_j\Phi^{\dagger}_k$
etc in the superpotential give rise to extra components of superfields that
do not appear in the expansion of chiral superfield. Thus $\theta\theta$ component
of the superpotential would not transform into itself plus a total derivative
and thereby would break SUSY.

As follows from Eq.~(\ref{ss23}) the first term in Eq.~(\ref{lag2}) leads to the
usual kinetic terms for the components of chiral superfields and $F^{\dagger}_i F_i$.
The term $[\Phi_i^{\dagger} \Phi_i]_{D}$ has mass dimension 4. Thereby high
products of left--handed and right--handed chiral superfields such as
$\Phi_i\Phi_j\Phi^{\dagger}_k$, $\Phi_i\Phi_j\Phi^{\dagger}_k \Phi^{\dagger}_l$
etc would result in couplings with negative mass dimension, i.e. non--renormalizable
interactions. Superpotential in Eq.~(\ref{lag2}) gives rise to mass terms,
Yukawa couplings and scalar potential. Combining Eqs.~(\ref{ss23}) and (\ref{ss20})
one obtains an explicit expression for the most general renormalizable
SUSY invariant Lagrangian in the Wess--Zumino model
\begin{eqnarray}
{\cal L}_{WZ} = \partial_{\mu}\phi^{\dagger}_i\partial^{\mu}\phi_i+
i \bar{\psi}_i \bar{\sigma}^{\mu}\partial_{\mu}\psi_i + F^{\dagger}_i F_i \qquad\qquad\qquad\nonumber\\[2mm]
\qquad\qquad\qquad + \left(\frac{\partial W(\phi_k)}{\partial \phi_i}F_i -
\frac{1}{2}\frac{\partial^2 W(\phi_k)}{\partial \phi_i\partial
\phi_j}\psi_i\psi_j + h.c.\right).
\label{lag4}
\end{eqnarray}
Note that in the Lagrangian (\ref{lag4}) there is no any kinetic terms for
the fields $F_i$. This means that the equations of motion for $F_i$ and
$F^{\dagger}_i$ reduce to algebraic equations, i.e.
\begin{eqnarray}
F^{\dagger}_i=-\frac{\partial W(\phi_k)}{\partial \phi_i}
\label{lag5}
\end{eqnarray}
Last equation shows that the fields $F_i$ are auxiliary fields which may
be eliminated. Using (\ref{lag5}) the Lagrangian becomes
\begin{equation}
{\cal L}_{WZ} = \partial_{\mu}\phi^{\dagger}_i\partial^{\mu}\phi_i+
i \bar{\psi}_i \bar{\sigma}^{\mu}\partial_{\mu}\psi_i - V(\phi_k)
-\frac{1}{2}\left(\frac{\partial^2 W(\phi_k)}{\partial \phi_i\partial
\phi_j}\psi_i\psi_j + h.c.\right),
\label{lag6}
\end{equation}
\begin{equation}
V(\phi_k)=\sum_i F^{\dagger}_i F_i = \sum_i \biggl|\frac{\partial W(\phi_k)}{\partial \phi_i}\biggr|^2\,.
\label{lag7}
\end{equation}
One can see that the tree-level scalar potential (\ref{lag7}) is positive definite.
Moreover the couplings that determine the interactions of scalar fields in
the potential (\ref{lag7}) are related to the Yukawa couplings..

The Lagrangian (\ref{lag2}) can be written in a much more elegant way
as an integral over the Grassmann variables $\theta$ and $\bar{\theta}$
(over superspace). The integration over Grassmann variables is defined
such that
\begin{equation}
\int\, d\theta^\alpha = \int d\bar{\theta}^{\dot{\alpha}}=0\,,\qquad
\int\, d\theta^\beta \theta^\alpha =
\int\, d\bar{\theta}^{\dot{\beta}}\bar{\theta}^{\dot{\alpha}} =
\delta_{\alpha\beta}\,.
\label{lag8}
\end{equation}
This allows an arbitrary function of $\theta$ and $\bar{\theta}$ to be
integrated , because for Grassmann variables we need never consider
powers higher than the first power of any component of $\theta$ or
$\bar{\theta}$. Volume elements in superspace are defined by
\begin{equation}
d^2\theta=-\frac{1}{4}\,d\theta^{\alpha}\,d\theta_{\alpha}\,,\qquad
d^2\bar{\theta}=-\frac{1}{4}\,d\bar{\theta}_{\dot{\alpha}}\,
d\bar{\theta}^{\dot{\alpha}}\,,\qquad d^4\theta=d^2\theta\, d^2\bar{\theta}
\label{lag9}
\end{equation}
It then follows from Eq.~(\ref{lag8}) that the non--zero integrals
over superspace are
\begin{equation}
\int d^2\theta\, \theta^2=\int d^2\bar{\theta}\, \bar{\theta}^2=1\,.
\label{lag10}
\end{equation}
The Lagrangian (\ref{lag2}) may now be written as
\begin{eqnarray}
{\cal L} & = &\int d^4\theta \sum_i \Phi_i^{\dagger}\Phi_i +
\biggl( \int d^2\theta\, W(\Phi_k) + h.c.\biggr)
\label{lag11}
\end{eqnarray}
because the superspace integrations project out $D$- and $F$-terms.

Using the supergraph techniques which are based on superspace
integrations the {\bf non--renormalization theorem} was proven.
This theorem may be stated as follows.

{\it The superpotential (for $N=1$ SUSY theory) is not renormalized,
except by finite amounts, in any order of perturbation theory, other
than by wave function renormalizations.}

The non--renormalization theorem derives from the observation that in
supergraph perturbation theory any radiative correction to the effective
action can be written as a single superspace integration $\int d^4 \theta$
over a product of quantities that are local in $\theta$ and $\bar{\theta}$
with no factors of superspace $\delta$--functions. The superpotential
term in (\ref{lag11}) is not of this form because it involves only
$\int d^2 \theta$. At the same time the $\Phi_i^{\dagger}\Phi_i$ terms
in the Lagrangian (\ref{lag11}) are renormalized resulting in wave
function renormalizations. Thus, any renormalization of masses and
coupling constants in SUSY models is due to wave function renormalization.

\subsection{SUSY QED}
Our next step is to construct gauge invariant SUSY Lagrangians.
It seems to be reasonable to start from the SUSY generalization
of the QED. In order to construct supersymmetric gauge field theory
we need to construct the field strength superfield and to couple
the vector superfield to the charged matter supermultiplets in a
gauge--invariant way. In the QED we apply partial derivatives to
vector field $V_{\mu}$ to get field strength tensor, i.e.
$V_{\mu\nu}=\partial_{\mu} V_{\nu}-\partial_{\nu} V_{\mu}$. In
SUSY QED one should use covariant derivatives for this purpose.
In the previous section we observed that the fields $\lambda$,
$\bar{\lambda}$, $V_{\mu\nu}$ and $D$ form an irreducible
representation of the SUSY algebra. Also all of these fields
are gauge invariant. This suggests that $\lambda$, $V_{\mu\nu}$
and $D$ may form components of the field strength superfield.
Then $\lambda_{\alpha}$ is the lowest--dimension component of
such superfield since $[\lambda_{\alpha}]=3/2$ while $
[V_{\mu\nu}]=[D]=2$. Since the mass dimensions of vector superfield
and covariant derivatives $D_{\alpha}$ and $\bar{D}_{\dot{\alpha}}$
are $[V]=0$, $[D_{\alpha}]=[\bar{D}_{\dot{\alpha}}]=1/2$
the required superfield can be obtained if we apply three
covariant derivatives to vector superfield, i.e.
\begin{equation}
W_{\alpha} = \bar{D}^2 D_{\alpha} V\,,
\label{lag12}
\end{equation}
where $\bar{D}^2=\bar{D}_{\dot{\alpha}} \bar{D}^{\dot{\alpha}}$.
From Eq.~(\ref{lag12}) it follows that $W_{\alpha}$ is a
spinor chiral superfield because
\begin{equation}
\bar{D}_{\dot{\beta}} W_{\alpha} = 0\,.
\label{lag13}
\end{equation}
Indeed, $\bar{D}_{\dot{\beta}}\bar{D}_{\dot{\gamma}}\bar{D}_{\dot{\sigma}}$
always give zero. Direct calculation gives
\begin{equation}
W_{\alpha}(y,\theta) = 4i\lambda_{\alpha}(y) - 
\theta_{\beta}\biggl[4\delta_{\alpha}^{\beta} D(y) + 2i(\sigma^{\mu}\bar{\sigma}^{\nu})_{\alpha}^{\beta} 
V_{\mu\nu}(y)\biggr] + 4\theta^2\sigma^{\mu}_{\alpha\dot{\alpha}} \partial_{\mu} \bar{\lambda}^{\dot{\alpha}}(y)\,.
\label{lag14}
\end{equation}
As before the spinor chiral superfield $W_{\alpha}$ is a function of $y$ and $\theta$ only.
$W^{\alpha} W_{\alpha}$ is a chiral scalar superfield whose lowest component is a scalar.
To construct the Lagrangian of SUSY QED we need to include $F$--component of $W^{\alpha} W_{\alpha}$
since it is invariant under Lorentz transformations, contains kinetic term for the $U(1)$ gauge
field $V_{\mu}$ and transforms as a total divergence under SUSY transformations.
A simple calculation yields
\begin{equation} 
\frac{1}{32} [W^{\alpha}W_{\alpha}]_{F} = -\frac{1}{4} V^{\mu \nu} V_{\mu \nu}
+i\lambda \sigma^{\mu}\partial_{\mu}\bar{\lambda} - 
\frac{i}{8} V^{\mu \nu} V^{\rho \sigma}\epsilon_{\mu \nu \rho \sigma } + \frac{1}{2}D^2\,.
\label{lag15}
\end{equation}
Eq.~(\ref{lag15}) should be considered as the supersymmetric generalization of the familiar
kinetic term $-\frac{1}{4} V^{\mu \nu} V_{\mu \nu}$ of the $U(1)$ gauge field.
The third term in Eq.~(\ref{lag15}) is a total divergence and does not affect the
equations of motion. The $D$--field is an auxiliary field which can be eliminated
using the equations of motion.

To go beyond a pure gauge theory we also need a SUSY version of the interaction of the
gauge field with charged matter. As in the usual QED we start from the kinetic terms
of charged matter which are contained in $[\Phi^{\dagger}_i\Phi_i]_D$. One can
expect that chiral superfields transform under the $U(1)$ gauge 
transformations like in the usual QED, i.e.
\begin{equation}
\Phi_i \rightarrow e^{-2i\,q_i\Lambda}\Phi_i\,, \qquad\qquad 
\Phi_i^{\dagger}\rightarrow \Phi_i^{\dagger} e^{2i\,q_i\Lambda^{\dagger}}\,,
\label{lag16}
\end{equation}
where $\Lambda $ is the scalar chiral superfield associated with the $U(1)$ gauge
transformation and $q_i$ are charges of matter superfields $\Phi_i$. Then gauge
transformation of vector superfield (\ref{ss29}) indicates that the combination
of superfields $\Phi^{\dagger}_i e^{2q_i V} \Phi_i$ is $U(1)$ gauge invariant.
Indeed
\begin{equation}
\Phi^{\dagger}_i e^{2q_i V} \Phi_i \rightarrow
\Phi_i^{\dagger} e^{2i\,q_i\Lambda^{\dagger}} e^{2q_i V + 2i\,q_i(\Lambda - \Lambda^{\dagger})}
e^{-2i\,q_i\Lambda}\Phi_i=\Phi^{\dagger}_i e^{2q_i V} \Phi_i\,.
\label{lag17}
\end{equation}
The combination of superfields (\ref{lag17}) is a real superfield, since $V$ is real.
Therefore the D--term of superfield (\ref{lag17}) yields a SUSY--invariant action.

In the Wess--Zumino gauge the exponential
\begin{equation}
e^{2q_i V}=1+2q_i V + 2q^2_i V^2 
\label{lag18}
\end{equation}
since $V^n=0$ if $n>2$. The leading term of the exponential gives $\Phi^{\dagger}_i\Phi_i$
The appearance of interaction terms proportional to $q$ and $q^2$ is also to be expected 
since in a SUSY theory there must also appear interactions of the gauge field with the
charged scalar particles. The combination of superfields (\ref{lag17}) can be expressed
in terms of the component fields of the superfields $\Phi_i$ and $V$, i.e.
\begin{equation}
\begin{array}{c}
\biggl[\Phi^{\dagger}_i e^{2q_i V} \Phi_i \biggr]_D =
(D_{\mu}\phi_i)^{\dagger}(D^{\mu}\phi_i)+
i \bar{\psi}_i \bar{\sigma}^{\mu}D_{\mu}\psi_i + F^{\dagger}_i F_i+\\[3mm]
i\sqrt{2} q_i (\phi_i^{\dagger}\psi_i\lambda-\phi_i\bar{\psi}_i\bar{\lambda})+q_i\phi_i^{\dagger}\phi_i D\,, 
\end{array}
\label{lag19}
\end{equation}
where $D^{\mu}=\partial^{\mu}+iq_i V^{\mu}$. From Eq.~(\ref{lag19}) one can see that 
partial derivatives $\partial^{\mu}$ in the expression for $\Phi^{\dagger}_i\Phi_i$
got replaced by $D^{\mu}$ providing an adequate description of fermionic and bosonic 
fields in the external field $V_{\mu}$. 

Putting Eqs.~(\ref{lag15}) and (\ref{lag19})
together yields the Lagrangian for the supersymmetric $U(1)$ gauge--invariant theory
\begin{eqnarray}
{\cal L}_{abel} & = & \frac{1}{32}\int d^2\theta
~W^{\alpha}W_{\alpha}  + \int d^2\theta  d^2\bar{\theta} ~\Phi_i^{\dagger} e^{2q_i V} \Phi_i \nonumber \\
&&+ \biggl(\int d^2\theta ~ W_{abel}(\Phi_k)  + h.c.\biggr)\,,
\label{lag20}
\end{eqnarray} 
where $W(\Phi_k)$ is a superpotential
\begin{eqnarray}
W_{abel}(\Phi_k)=\frac{1}{2}m_{ij}\Phi_i \Phi_j + \frac{1}{3} y_{ijk} \Phi_i \Phi_j \Phi_k\,, 
\label{lag21}
\end{eqnarray} 
which is invariant under the Abelian $U(1)$ gauge group transformations. In other words the total 
charge of each term in Eq.~(\ref{lag21}) has to vanish to preserve gauge invariance. Here we assume that
all chiral superfields $\Phi_i$ carry non--zero $U(1)$ charges $q_i$. As a consequence all
linear terms in the superpotential (\ref{lag21}) are forbidden by the $U(1)$ gauge symmetry.
The first term in the Lagrangian (\ref{lag20}) contains kinetic terms of the $U(1)$ gauge field
and its superpartner (gaugino). The second term in (\ref{lag20}) includes all kinetic terms
of bosonic and fermionic components of chiral superfields $\Phi_i$ as well as the interactions
of charged fermions (bosons) with the $U(1)$ gauge boson (and gaugino). The last term in (\ref{lag20})
gives rise to the Yukawa interactions of bosonic and fermionic components of chiral superfields $\Phi_i$.

The simplest version of SUSY QED should contain at least one charged massive fermion, such as an electron.
To describe this massive field we need to include both its left and right chiral components. Thus we have
to employ two left--handed chiral superfields. One of them, $S$, contains left--handed electron (as $\psi_S$)
and its superpartner, the left--handed selectron (as $\phi_S$). Another chiral superfield, $T^{\dagger}$,
involves the right--handed electron (as $\bar{\psi}_T$) and its SUSY partner the right--handed selectron
(as $\phi^{\dagger}_T$). The Lagrangian of the corresponding SUSY generalization of QED looks as follows:
\begin{eqnarray}
{\cal L}_{SUSY \, QED} & = & \frac{1}{32}\int d^2\theta ~W^{\alpha}W_{\alpha} + 
\int d^4\theta ~(S^{\dagger} e^{2qV} S + T^+ e^{-2qV} T) \nonumber \\
&+& \int d^2\theta ~m~S\, T  + \int d^2\bar{\theta} ~m~S^{\dagger}\, T^{\dagger}\,.
\label{lag22}
\end{eqnarray}
Using the earlier results (\ref{lag15}), (\ref{lag19}) one can write the Lagrangian of SUSY QED in
terms of the components of superfields $S,\,T$ and $V$
\begin{eqnarray}
{\cal L}_{SUSY \, QED} & = & (D_{\mu}\phi_S)^{\dagger}_i(D^{\mu}\phi_S) + 
(D_{\mu}\phi_T^{\dagger})(D^{\mu}\phi_T^{\dagger})^{\dagger}+
i \bar{\psi}_S \bar{\sigma}^{\mu}D_{\mu}\psi_S \nonumber\\[2mm]
&+& i \bar{\psi}_T \bar{\sigma}^{\mu}D^{\dagger}_{\mu}\psi_T + F^{\dagger}_S F_S + F^{\dagger}_T F_T +
i\sqrt{2} q (\phi_S^{\dagger}\psi_S - \phi_T^{\dagger}\psi_T)\lambda\nonumber\\[2mm]
&-&i\sqrt{2} q (\phi_S\bar{\psi}_S-\phi_T\bar{\psi}_T)\bar{\lambda} + q (\phi_S^{\dagger}\phi_S - 
\phi_T^{\dagger}\phi_T) D \label{lag23}\\[2mm]
&+& m (\phi_S F_T + \phi_T F_S + \phi^{\dagger}_S F^{\dagger}_T + \phi^{\dagger}_T F^{\dagger}_S 
- \psi_S\psi_T - \bar{\psi}_S\bar{\psi}_T)\nonumber \\[2mm]  
&-&\frac{1}{4} V^{\mu \nu} V_{\mu \nu} + i\lambda \sigma^{\mu}\partial_{\mu}\bar{\lambda} 
+ \frac{1}{2}D^2\,.\nonumber
\end{eqnarray}
The $F$--components of superfields $S$ and $T$ ($F_S$ and $F_T$) and D--component of vector
superfield $V$ are auxiliary fields since Lagrangian (\ref{lag23}) does not contain
their derivatives. Using field equations 
\begin{eqnarray}
F_S+m\phi_T^{\dagger}=F_T+m\phi_S^{\dagger}=0 \nonumber\\
D+q (\phi_S^{\dagger}\phi_S - \phi_T^{\dagger}\phi_T)=0
\label{lag24}
\end{eqnarray} 
$F_S$, $F_T$ and $D$ can be eliminated. The resulting Lagrangian can be presented in 
the following form:
\begin{eqnarray}
{\cal L}_{SUSY \, QED} & = & (D_{\mu}\phi_S)^{\dagger}_i(D^{\mu}\phi_S) - m^2 \phi_S^{\dagger} \phi_S
+(D_{\mu}\phi_T^{\dagger})(D^{\mu}\phi_T^{\dagger})^{\dagger} - m^2 \phi_T^{\dagger} \phi_T \nonumber\\[2mm]
&+&i \bar{\psi}_S \bar{\sigma}^{\mu}D_{\mu}\psi_S 
+i\bar{\psi}_T \bar{\sigma}^{\mu}D^{\dagger}_{\mu}\psi_T - m(\psi_S\psi_T + \bar{\psi}_S\bar{\psi}_T) \nonumber\\[2mm] 
&+& i\sqrt{2} q (\phi_S^{\dagger}\psi_S - \phi_T^{\dagger}\psi_T)
-i\sqrt{2} q (\phi_S\bar{\psi}_S-\phi_T\bar{\psi}_T)\bar{\lambda} \\[2mm]
&-& V_{QED}(\phi_S, \phi_T) - \frac{1}{4} V^{\mu \nu} V_{\mu \nu} + i\lambda \sigma^{\mu}\partial_{\mu}\bar{\lambda}
\,,\nonumber
\label{lag25}
\end{eqnarray}  
where $V_{QED}(\phi_S, \phi_T)$ is a quartic part of the scalar potential
\begin{eqnarray}
V_{QED}(\phi_S, \phi_T)= \frac{1}{2} D^2 = \frac{q^2}{2} (\phi_S^{\dagger}\phi_S - \phi_T^{\dagger}\phi_T)^2\,.
\label{lag26}
\end{eqnarray}
The first two lines in Eq.~(\ref{lag25}) represent the Lagrangian of the QED model that contains 
massive electron and two massive scalar fields with the same electric charge. Supersymmetry forces
these fields to have the same mass. SUSY also results in the presence of massless Majorana fermion
(gaugino/photino) in the particle spectrum. This fermion interacts with the electron and scalar fields.
Supersymmetry ensures that the corresponding Yukawa couplings are proportional to the electric charge
that electron and scalar fields have. SUSY also determines the form of the quartic interactions
in the scalar potential (\ref{lag26}). The corresponding self--couplings of the scalars are set by 
the $U(1)$ electric charge as well.

\subsection{Supersymmetric non--abelian gauge theories}
If SUSY is realized in Nature, it is certainly at an energy scale that is higher than that of the 
EW scale. It is therefore essential to have a supersymmetric extension not only of the $U(1)$
abelian gauge invariance but also of the non--abelian gauge invariance that occurs in the SM
and Grand Unified Theories (GUTs). In the non--abelian theories there is a set of gauge fields
that form a representation of a non--abelian group $G$. Therefore we need to introduce a set of
vector superfields that transform under the non--abelian gauge group transformations. Thus
we replace vector superfield $V$ by $V^a T^a$ where
\begin{eqnarray}
V^a(x, \theta, \bar \theta) & = & C^a(x) + i\theta \chi^a (x)
-i\bar{\theta}\bar{\chi}^a (x)  + \frac{i}{2} \theta \theta [M^a (x) + iN^a (x)] \nonumber\\
&&-\frac{i}{2} \bar{\theta}\bar{\theta}[M^a (x)-iN^a (x)] -
\theta\sigma^{\mu}\bar{\theta}V^a_{\mu}(x)\\
&&+i\theta\theta\bar{\theta}\biggr[\bar{\lambda}^a(x)
+\frac{i}{2}\bar{\sigma}^{\mu}\partial_{\mu}\chi^a (x)\biggl]
- i\bar{\theta}\bar{\theta}\theta \biggr[\lambda^a(x) +
\frac{i}{2}\sigma^{\mu}\partial_{\mu}\bar{\chi}^a(x)\biggl]
\nonumber\\ 
&&+\frac{1}{2} \theta \theta \bar{\theta}\bar{\theta}\biggr[D^a (x) -
\frac{1}{2}\partial_{\mu}\partial^{\mu} C^a (x)\biggl]
\nonumber\,.
\label{lag27}
\end{eqnarray}
In SUSY QCD $V^a_{\mu}(x)$ and $\lambda^a(x)$ should be associated with the gluons
and their superpartners (gluinos) that form adjoint representation of $SU(3)$. 
Matrices $T^a$ represent the generators of non--abelian group $G$ that obey algebra
\begin{eqnarray}
\biggl[T^a,\,T^b\biggr] = i f^{abc} T^c\,,
\label{lag28}
\end{eqnarray}
where $f^{abc}$ are the totally antisymmetric structure constants of $G$.

One can expect that the multiplets of chiral superfields $\Phi_i$ transform 
under the non--abelian gauge group transformations as follows
\begin{equation}
\Phi_i \rightarrow e^{-2i\,g T^a \Lambda^a}\Phi_i\,, \qquad\qquad
\Phi_i^{\dagger}\rightarrow \Phi_i^{\dagger} e^{2i\,g T^a \Lambda^{a\dagger}}\,,
\label{lag29}
\end{equation}  
where $\Lambda^a $ are chiral superfields associated with the non--abelian gauge 
transformations and $g$ is the corresponding gauge coupling.
As in the abelian case the kinetic terms of bosonic and fermionic components of
chiral superfields $\Phi_i$ can originate from 
\begin{equation}
\biggl[\Phi^{\dagger}_i e^{2g\, T^a V^a} \Phi_i \biggr]_D\,.
\label{lag30}
\end{equation}
The combination of superfields (\ref{lag30}) yields interaction of gauge 
and matter fields which is gauge invariant. The gauge invariance of the 
above combination follows provided that the gauge--transformed vector
superfields $V'^a$ satisfy 
\begin{equation}
e^{2V'} = e^{-2i\,\Lambda^{\dagger}} e^{2V} e^{2i\,\Lambda}\,,
\label{lag31}
\end{equation}
where $V=g V^a T^a$ and $\Lambda=g\Lambda^a T^a$. 

Eq.~(\ref{lag31}) can be considered as the non--abelian generalization of Eq.~(\ref{ss29}).
At the same time Eq.~(\ref{lag31}) does not get reduced to Eq.~(\ref{ss29}) 
because the generators of non--abelian group do not commute. Under an 
infinitesimal gauge transformation 
\begin{eqnarray}
V'& = & V + \delta V\,,\nonumber\\
\delta V & = & i (\Lambda-\Lambda^{\dagger})+i\biggl[V,\,\Lambda+\Lambda^{\dagger}\biggr]+
\frac{i}{3}\biggl[V,\biggl[V,\,\Lambda-\Lambda^{\dagger}\biggr]\biggr]+...\,. 
\label{lag32}
\end{eqnarray}
The right hand side of Eq.~(\ref{lag32}) contains infinite tower of higher commutators.
It is easy to see that the first two terms in Eq.~(\ref{lag32}) generate the
familiar gauge transformation of the non--abelian vector potential 
\begin{eqnarray}
V'^{a}_{\mu}=V^a_{\mu}+\partial_{\mu}(\omega^a+\omega^{a\dagger})+
g f^{abc}(\omega^b+\omega^{b\dagger})V^c_{\mu}\,,
\label{lag33}   
\end{eqnarray}
where $\omega^a$ is the scalar component of $\Lambda^a$.
Eq.~(\ref{lag32}) indicates that in the non--abelian case it is still possible 
to arrange $\Lambda$ and $\Lambda^{\dagger}$ such that $C^a$, $\chi^a$, $M^a$ and $N^a$
components of vector superfield $V^a$ vanish giving Wess--Zumino gauge. In this gauge
\begin{eqnarray}
V^a(x, \theta, \bar \theta) =  \theta\sigma^{\mu}\bar{\theta}V^a_{\mu}(x)
+i\theta\theta\bar{\theta}\bar{\lambda}^a(x)
- i\bar{\theta}\bar{\theta}\theta \lambda^a(x) 
+\frac{1}{2} \theta \theta \bar{\theta}\bar{\theta} D^a (x)\,.
\label{lag34}   
\end{eqnarray}

Now we have to construct the non--abelian strength superfield, analogous to (\ref{lag12}).
The $W_{\alpha}$ defined by Eq.~(\ref{lag12}) is not gauge invariant in the non--abelian
case. The non--abelian gauge transformation (\ref{lag31}) suggests that we need to use
$e^{2V}$ instead of $V$. If we define 
\begin{eqnarray}
W_{\alpha} =  \frac{1}{2g} \bar{D}^2 e^{-2V} D_{\alpha} e^{2V}\,, \qquad
W^{\dagger}_{\dot{\alpha}} = - \frac{1}{2g} D^2 e^{2V} \bar{D}_{\dot{\alpha}} e^{-2V},
\label{lag35}
\end{eqnarray}
then $W_{\alpha}$ and $W^{\dagger}_{\dot{\alpha}}$ transform under the non--abelian transformations
in the following way:
\begin{eqnarray}
W_{\alpha} \rightarrow e^{-2i\,\Lambda} W_{\alpha} e^{2i\,\Lambda}\,,\qquad
W^{\dagger}_{\dot{\alpha}} \rightarrow 
e^{-2i\,\Lambda^{\dagger}} W^{\dagger}_{\dot{\alpha}} e^{2i\,\Lambda^{\dagger}}\,,
\label{lag36} 
\end{eqnarray}
where $W_{\alpha}=W^a_{\alpha} T^a$ and $W^{\dagger}_{\dot{\alpha}}=W^{a\dagger}_{\dot{\alpha}} T^a$.
Eqs.~(\ref{lag36}) suggest that 
$$
\mbox{Tr} W^{\alpha} W_{\alpha}=\frac{1}{2} W^{a\alpha} W^a_{\alpha}\,,\qquad  
\mbox{Tr} W^{\dagger}_{\dot{\alpha}} W^{\dot{\alpha}\dagger}=
\frac{1}{2} W^{a\dagger}_{\dot{\alpha}} W^{a\dot{\alpha}\dagger}
$$
are gauge invariant.
This is completely analogous to the non--supersymmetric case where the field
strength tensor $F_{\mu\nu}$ itself is invariant in the abelian case but
in the non--abelian case only the trace $\mbox{Tr} F^{\mu\nu} F_{\mu\nu}$ is
gauge invariant with $F_{\mu\nu}=F^a_{\mu\nu} T^a$\,.
Expanding in powers of $V$ gives
\begin{eqnarray}
W_{\alpha} =  \frac{1}{g} \bar{D}^2 (D_{\alpha} V + [D_{\alpha} V,\,V]+...)\,.
%\bar{W}_{\dot{\alpha}} = \frac{1}{g} D^2 (\bar{D}_{\dot{\alpha}} V - [\bar{D}_{\dot{\alpha}} V,\,V]+...)\,.
\label{lag37}
\end{eqnarray}
In the Wess--Zumino gauge only the first two terms survive. Thus we have
\begin{eqnarray}
W^a_{\alpha}= \bar{D}^2 D_{\alpha} V^a + i g f^{abc} \bar{D}^2 (D_{\alpha} V^b) V^c\,.
\label{lag38}
\end{eqnarray}
From Eq.~(\ref{lag38}) one can see that $W^a_{\alpha}$ reduces to Eq.~(\ref{lag12})
when $f^{abc}\to 0$ that corresponds to the abelian limit. 

In terms of the component fields of the superfields $V^a$ we get
\begin{eqnarray}
W^a_{\alpha}(y,\theta) & = & 4i\lambda^a_{\alpha}(y) +
\theta_{\beta}\biggl[4\delta_{\alpha}^{\beta} D^a(y) + 2i(\sigma^{\mu}\bar{\sigma}^{\nu})_{\alpha}^{\beta}
V^a_{\mu\nu}(y)\biggr] \nonumber \\[2mm] 
& + & 4\theta^2\sigma^{\mu}_{\alpha\dot{\alpha}} D_{\mu} \bar{\lambda}^{a\dot{\alpha}}(y) ,
\label{lag39}
\end{eqnarray}
where 
\begin{eqnarray}
V^a_{\mu\nu}=\partial_{\mu} V^a_{\nu} - \partial_{\nu} V^a_{\mu} - g f^{abc} V^b_{\mu} V^c_{\nu}\,,\qquad
D_{\mu} \bar{\lambda}^{a\dot{\alpha}} = \partial_{\mu} \bar{\lambda}^{a\dot{\alpha}}
-g f^{abc} V^b_{\mu} \bar{\lambda}^{c\dot{\alpha}}\,.
\label{lag40} 
\end{eqnarray}
From the definition of the non--abelian strength superfield (\ref{lag35}) and Eqs.~(\ref{lag37})--(\ref{lag39})
it follows that as before $W^a_{\alpha}(y,\theta)$ is a set of spinor chiral superfields.
Then $W^{a\alpha} W^a_{\alpha}$ and $W^{a\dagger}_{\dot{\alpha}} W^{a\dot{\alpha}\dagger}$
are scalar chiral superfields which are invariant under non--abelian gauge transformation.
Therefore the $F$--components of these chiral superfields can be used for the 
construction of the Lagrangians of SUSY QCD and other supersymmetric non--abelian gauge theories.
In the Wess--Zumino gauge direct calculation gives
\begin{eqnarray}
\frac{1}{64} \biggl[(W^{a\alpha} W^a_{\alpha}) + 
(W^{a\dagger}_{\dot{\alpha}} W^{a\dot{\alpha}\dagger})\biggr]_F=
-\frac{1}{4} V^{a}_{\mu \nu} V^{a\mu \nu}+i\lambda^a \sigma^{\mu}D_{\mu}\bar{\lambda}^{a} 
+\frac{1}{2} D^a D^a\,.
\label{lag41} 
\end{eqnarray}
Eq.~(\ref{lag41}) is the SUSY generalization of the term $-\frac{1}{4}F^a_{\mu\nu} F^{a\mu\nu}$
in the non--abelian gauge theories.

Now we can specify the full Lagrangian of SUSY model based on the non--abelian gauge group $G$.
It can be presented in the following form
\begin{eqnarray}
{\cal L}_{G} & = & \frac{1}{32}\int d^2\theta~\mbox{Tr} W^{\alpha} W_{\alpha}+
\frac{1}{32}\int d^2\bar{\theta}~\mbox{Tr} W^{\dagger\alpha} W^{\dagger}_{\alpha} \nonumber\\[2mm]
&+&\int d^2\theta  d^2\bar{\theta} ~\Phi_i^+ e^{2 V} \Phi_i + 
\biggl(\int d^2\theta ~ W_{G}(\Phi_k)  + h.c.\biggr)\,,
\label{lag42}
\end{eqnarray} 
where $W_{G}(\Phi_k)$ is a superpotential which is required to be invariant under the action
of non--abelian group $G$ (as well as no more than cubic in the chiral superfields $\Phi_k$).
The first two terms in Eq.~(\ref{lag42}) contain the kinetic terms of the non--abelian gauge
fields (gluons) and their superpartners (gauginos) as well as their self interactions due to 
the non--abelian nature of the gauge group $G$. Thus in contrast with SUSY QED the first two 
terms in Eq.~(\ref{lag42}) leads to a non--abelian gaugino--gaugino-gauge boson interaction
through the term $\lambda^a \sigma^{\mu}D_{\mu}\bar{\lambda}^{a}$. The third term in 
Eq.~(\ref{lag42}) provides the kinetic terms for the bosonic and fermionic components of 
chiral superfields (squarks and quarks) as well as the gauge interactions of these states
with the non--abelian gauge fields and gauginos. Finally, last terms in Eq.~(\ref{lag42})
result in the Yukawa interactions of the bosonic and fermionic components of $\Phi_k$.
In general there might be several chiral supermultiplets $\Phi_k$ which form 
different representations of the non--abelian group $G$. Then for each $\Phi_i$
one has to use for $V=g V^a T^a$ the matrix constructed with the representation $T^a_i$
appropriate to $\Phi_i$.

In terms of the component fields of superfields the Lagrangian (\ref{lag42}) 
takes the form
\begin{eqnarray}
{\cal L}_{G} & = & -\frac{1}{4} V^{a}_{\mu \nu} V^{a\mu \nu}+i\lambda^a \sigma^{\mu}D_{\mu}\bar{\lambda}^{a}
+\frac{1}{2} D^a D^a +(D_{\mu}\phi_i)^{\dagger}(D^{\mu}\phi_i) \nonumber \\[1mm]
&+& i \bar{\psi}_i \bar{\sigma}^{\mu}D_{\mu}\psi_i + F^{\dagger}_i F_i
+ i\sqrt{2} g\, (\phi_i^{\dagger} T^a \lambda^a \psi_i -\phi_i\bar{\psi}_i T^a \bar{\lambda}^a) \label{lag43}\\[1mm]
&+& g\, \phi_i^{\dagger}T^a D^a \phi_i  + \left(\frac{\partial W(\phi_k)}{\partial \phi_i}F_i -
\frac{1}{2}\frac{\partial^2 W(\phi_k)}{\partial \phi_i\partial \phi_j}\psi_i\psi_j + h.c.\right)\,,\nonumber
\end{eqnarray}
where 
$$
D_{\mu}\phi_i = \partial_{\mu}\phi_i+ig\, T^a V^a_{\mu} \phi_i\,,\qquad 
D_{\mu}\psi_i = \partial_{\mu}\psi_i+ig\, T^a V^a_{\mu} \psi_i\,.
$$
Again, the Lagrangian (\ref{lag43}) does not have any kinetic terms for the auxiliary fields 
$F_i$ and $D^a$. Thereby these fields can be eliminated using their equations of 
motion, i.e.
\begin{eqnarray}
F^{\dagger}_i=-\frac{\partial W(\phi_k)}{\partial \phi_i}\,,\qquad\qquad 
D^a=-\sum_i g\,\phi_i^{\dagger}T^a \phi_i\,.
\label{lag45}
\end{eqnarray}
Substituting $F^{\dagger}_i$ and $D^a$ back into Eq.~(\ref{lag43}) we obtain 
the resulting Lagrangian
\begin{eqnarray}
{\cal L}_{G} & = & -\frac{1}{4} V^{a}_{\mu \nu} V^{a\mu \nu}+i\lambda^a \sigma^{\mu}D_{\mu}\bar{\lambda}^{a}
+(D_{\mu}\phi_i)^{\dagger}(D^{\mu}\phi_i) \nonumber \\[2mm]
&+& i\bar{\psi}_i\bar{\sigma}^{\mu}D_{\mu}\psi_i +
i\sqrt{2}g\,(\phi_i^{\dagger} T^a\lambda^a\psi_i - \phi_i\bar{\psi}_i T^a \bar{\lambda}^a) \label{lag46} \\[2mm] 
&-&\left(\frac{1}{2}\frac{\partial^2 W(\phi_k)}{\partial \phi_i\partial \phi_j}\psi_i\psi_j + h.c.\right)
-V(\phi_k)\,,\nonumber
\end{eqnarray}
where~ the~ full~ scalar~ potential~ is~ the~ sum~ of~ two~ contributions~ from~ 
the~ $F$--terms and $D$--terms
\begin{eqnarray}
V(\phi_k) = \sum_i F^{\dagger}_i F_i + \sum_a \frac{1}{2} (D^a)^2 =
\sum_i \left|\frac{\partial W(\phi_k)}{\partial \phi_i}\right|^2+
\frac{g^2}{2}\sum_a \biggl(\sum_i \phi_i^{\dagger} T^a \phi_i \biggr)^2. 
\label{lag47}
\end{eqnarray}
From Eq.~(\ref{lag47}) it becomes obvious that the scalar potential in SUSY models is positive 
definite. It is completely defined by the superpotential and gauge interactions. 

Thus the form of the SUSY Lagrangian is practically fixed by symmetry requirements.
The only freedom is the field content, the values of the gauge and Yukawa couplings 
and the mass parameters in the superpotential.

\newpage
\section{The minimal SUSY model}
SUSY algebra implies that each SUSY multiplet must have equal number of bosonic
and fermionic degrees of freedom. As a result the simplest supersymmetric extensions
of the SM should contain scalar degrees of freedom associated with the left--handed
and right--handed SM fermions, i.e. left--handed and right--handed squarks and
sleptons. These models must also include the fermionic partners of the SM gauge bosons
(gauginos) and Higgs bosons (Higgsinos). 

In the SM one Higgs doublet $H$ is used to generate masses for up- and down--type 
quarks and charged leptons. These masses are induced by means of the Yukawa interactions 
of quarks and leptons with the Higgs fields. More precisely the masses of the down--type 
quarks and charged leptons are generated by the Higgs doublet itself whereas the 
conjugated Higgs doublet $i\sigma_2 H^{\dagger}$ gives rise to the masses of up--type 
quarks. The results of the previous section indicate that in SUSY models the Higgs--fermion 
Yukawa interactions can orginate from the superpotential only. Since $W(\Phi_k)$ is an 
analytic function of the chiral superfields it can not involve any conjugate superfield.
Thus we have no other choice than to introduce a second Higgs doublet $H_2$ with the 
opposite hypercharge which gives masses to the up--type quarks. The presence of the second 
Higgs doublet also ensures the cancellation of anomalies.

Thus the minimal supersymmetric standard model (MSSM) includes the following set of 
chiral superfields
\begin{equation} 
\begin{array}{ll}
Q_a=(u_a,\,d_a)\sim\left(3,\,2,\,\frac{1}{6}\right)\,,\qquad 
& u^c_a\sim\left(\bar{3},\,1,\,-\frac{2}{3}\right)\,,\\[2mm] 
L_a=(\nu_a,\,e_a)\sim\left(1,\,2,\,-\frac{1}{2}\right)\,,\qquad 
& d^c_a\sim\left(\bar{3},\,1,\,\frac{1}{3}\right)\,,\\[2mm]
H_{1}=(H^0_{1},\,H^{-}_{1i})\sim \left(1,\,2,\,-\frac{1}{2}\right)\,,\qquad 
& e^c_a\sim\left(1,\,1,\,1\right)\,,\\[2mm]
H_{2}=(H^{+}_{2},\,H^{0}_{2})\sim \left(1,\,2,\,\frac{1}{2}\right)\,, &
\end{array}
\label{mssm1}
\end{equation}
where the first and second quantities in the brackets are the $SU(3)_C$ and
$SU(2)_W$ representations, the third quantity in the brackets is the $U(1)_Y$ 
hypercharge, while $a$ is a family index that runs from 1 to 3.
Here $Q_a$ and $L_a$ contain the doublets of left--handed quark and lepton
superfields, $e^c_a$, $u^c_a$ and $d^c_a$ are associated with the right--handed
lepton, up-- and down--type quark superfields respectively whereas
$H_1$ and $H_2$ involve the doublets of Higgs superfields.
In Eq.~(\ref{mssm1}) and further we omit all isospin and colour indexes
related to $SU(2)_W$ and $SU(3)_C$ gauge interactions. 

In addition to Higgs, quark and lepton chiral superfields the MSSM includes 
three vector supermultiplets 
\begin{equation}
V_1\sim (1,\,1,\,0)\,,\qquad V_2\sim (1,\, 3,\,0)\,,\qquad V_3\sim (8,\,1,\,0)\,,
\label{mssm2}
\end{equation}
which are associated with $U(1)_Y$, $SU(3)_C$ and $SU(2)_W$ interactions.
$V_1$ is an abelian vector superfield that contains $U(1)_Y$ gauge field
and its superpartner which is called bino. $V_2$ involves a triplet of
$SU(2)_W$ gauge bosons and their superpartners (winos). $V_3$ includes
octet of gluons and octet of their superpartners (gluinos).

In order to reproduce the Higgs--fermion Yukawa interactions that induce
the masses of all quarks and charged leptons in the SM we need to 
include in the MSSM Lagrangian the following sum of the products of  
chiral superfields mentioned above 
\begin{equation}
W_{MSSM} = y^U_{ab}Q_a u^c_bH_2 + y^D_{ab} Q_a d^c_b H_1 + 
y^L_{ab} L_a e^c_b H_1 + \mu H_1 H_2\,,
\label{mssm3}
\end{equation}
where $a$ and $b$ are family indices. In Eq.~(\ref{mssm3}) the Yukawa couplings 
$y^U_{ab}$, $y^D_{ab}$ and $y^L_{ab}$ are dimensionless $3\times 3$ matrices in family 
space that determine the masses of quarks and charged leptons as well as the
phase of the CKM matrix. Here we also iclude a term $\mu H_1 H_2$ which is not 
present in the Lagrangian of the SM. It gives rise to the masses of the superpartners 
of Higgs bosons (Higgsinos). The $\mu$ term, as it is traditionally called,
can be written as $\mu (H_1)_{\alpha} (H_{2})_{\beta}\varepsilon^{\alpha\beta}$,
where $\varepsilon^{\alpha\beta}$ is used to tie together $SU(2)_W$ weak isospin 
indices $\alpha,\beta=1,2$ in a gauge invariant way. 

Eq.~(\ref{mssm3}) defines the simplest superpotential that the minimal SUSY model
can have. However there are extra terms that one can write which are gauge invariant
and analytic in the chiral superfields. These additional terms are given by
\begin{equation}
W_{NR}=\lambda^L_{abd}L_a L_b e_d + \lambda^{L\prime}_{abd}L_a Q_b d_d +\mu'_a L_a H_2+
\lambda^B_{abd} u_a^c d_b^c d_d^c,
\label{mssm4}             
\end{equation}
The terms in $W_{NR}$ violate either lepton or baryon number resulting in rapid 
proton decay. The most general renormalizable gauge invariant superpotential
of the simplest SUSY extension of the SM is a sum of Eqs.~(\ref{mssm3}) and 
(\ref{mssm4}), i.e. $W=W_{MSSM}+W_{NR}$. The terms given by Eq.~(\ref{mssm4})
are absent in the SM. The inclusion of such terms in the Lagrangian of the SM
would violate Lorentz invariance. Since B-- and L--violating processes have not
been observed in Nature, the terms in $W_{NR}$ must be very suppressed.

The baryon and lepton number violating processes in the MSSM can be suppressed by 
postulating the invariance of the Lagrangian under $R$--parity transformations ($P_R$)
or equivalently matter parity transformations ($P_M$)
\begin{equation}
P_R =(-1)^{3(B-L)+2s}\,,\qquad P_M=(-1)^{3(B-L)}\,,
\label{mssm5}
\end{equation}
where $s$ is the spin of the particle. It is easy to check that the quark and
lepton supermultiplets have $P_M=-1$, while the Higgs and vector supermultiplets
have $P_M=+1$. Matter parity forbids all terms in $W_{NR}$. This symmetry
commutes with SUSY, as all component fields of a given supermultiplet have 
the same matter parity. The advantage of matter parity is that it can in 
principle be an exact and fundamental symmetry whereas $B$ and $L$ themselves
cannot, since they are known to be violated by non--perturbative electroweak
effects. Indeed, matter parity can originate from the continuous $U(1)_{B-L}$
gauge symmetry that satisfies anomaly cancellation conditions. $P_M$ can
survive as an exactly conserved discrete remnant subgroup of $U(1)_{B-L}$.
Although matter parity forbids all renormalizable interactions which 
violate $B$ and $L$ in the MSSM one may expect that baryon and/or lepton number 
violation can occur in tiny amounts due to the non--renormalizable terms 
in the Lagrangian.

Matter parity conservation and $R$--parity conservation are equivalent, since 
the product of $(-1)^{2s}$ for the particles involved in any interaction vertex
in a theory, which conserves angular momentum, is always equal to $+1$.
At the same time particles within the same supermultiplet do not have the same 
$R$--parity and there is no any physical principle behind it. Due to the
matter parity it secretly does commute with SUSY. Nevertheless, the $R$--parity
assignment is very useful for phenomenology because all of the SM particles and the
Higgs bosons have even $R$--parity while all of the squarks, sleptons, gauginos and 
higgsinos have odd $R$--parity. The $R$--parity odd particles are known as 
"SUSY particles" or "sparticles". Since in the conventional MSSM $R$--parity
is conserved there can not be any mixing between states with $P_R=+1$ and
$P_R=-1$. Furthermore, every interaction vertex in the MSSM contains an
even number of $P_R=-1$ states. This has three important phenomenological
consequences:
\begin{itemize}
\item The lightest supersymmetric particle (LSP) must be absolutely stable and 
can play the role of non--baryonic dark matter. In most supersymmetric scenarios 
the LSP is the lightest neutralino which is a mixture of Higgsinos and gauginos. 
Since the lightest neutralino is a heavy weakly interacting particle it explains
well the large scale structure of the Universe and can provide the correct relic
abundance of dark matter if its mass is of the order of the EW scale.
\item In collider experiments sparticles can only be created in pairs.
\item Each sparticle must eventually decay into a final state that contains
an odd number of LSPs (usually just one). Since stable lightest neutralinos can 
not be detected directly their signature would be missing energy and transverse 
momentum in the final state. 
\end{itemize}

The form of the MSSM superpotential (\ref{mssm3}) can be simplified 
substantially. Since the top quark is the heaviest fermion one can expect 
that only $y^U_{33}=h_t$ is important while all other Yukawa couplings are 
negligibly small and can be ignored in the first approximation. Then 
MSSM superpotential takes the form
\begin{equation}
W_{MSSM} \simeq \mu H_1 H_2+h_t Q t^c H_2 \,,
\label{mssm6}
\end{equation}
where $Q=Q_3$ and $t^c=u^c_3$. The structure of SUSY Lagrangians discussed
in the previous section implies that the Yukawa coupling $h_t$ determines 
not only Higgs--quark--quark interaction but also squark--Higgsino--quark 
interaction. Moreover superpotential (\ref{mssm6}) also induces quartic 
interactions in the scalar potential
\begin{equation}
V_{F} = \sum_i F^{\dagger}_i F_i = h_t^2 |H_2 Q|^2 + h_t^2 |H_2|^2 |t^c|^2 +
|\mu H_1 + h_t Q t^c|^2\,.
\label{mssm7}  
\end{equation}
As one can see from Eq.~(\ref{mssm7}) the strength of these interactions is
set by $h_t^2$. Thus~~ Higgs--quark--quark,~~ squark--Higgsino--quark,~~
$(\mbox{squark})^2$ $(\mbox{Higgs})^2$~ and\\ 
(squark$)^4$ interactions
considered above are required by SUSY to have the same strength $h_t$.
This illustrates the remarkable economy of supersymmetry: there are many
interactions determined by only a single parameter. The SUSY relationships
between the dimensionless couplings provide an exact cancellation of 
quadratic divergences within supersymmetric models that allows to stabilise 
the mass hierarchy \cite{2}-\cite{5}. 

The cancellation of quadratic divergences in the MSSM is caused by
sparticles. Indeed, SUSY ensures that for each fermion loop there are
diagrams with SUSY partners (gauge bosons and/or scalars) in the loop. 
Due to the SUSY relationships between the dimensionless couplings
the contribution from boson loops cancels those from the fermion ones.
Thus the presence of sparticles at low energies play an extremely 
important role. However SUSY predicts that bosons and fermions from 
one SUSY multiplet have to be degenerate. Because superpartners of
quarks and leptons have not been observed yet, supersymmetry must be broken,
i.e. all sparticles have to be heavy. At the same time the supersymmetry 
breaking couplings should not spoil the cancellation of quadratic divergences. 
Terms fulfilling these requirements are called soft SUSY breaking terms. 
The Lagrangian of SUSY models based on the softly broken supersymmetry 
can be written as
\begin{equation}
{\cal L}={\cal L}_{SUSY}+{\cal L}_{soft}\,.
\label{mssm8}
\end{equation}

The part of the MSSM Lagrangian which is invariant under SUSY transformations
is given by
\begin{eqnarray}
{\cal L}^{MSSM}_{SUSY} = \sum_{A=SU(3),SU(2),U(1)}
\frac{1}{32}\left(\int d^2\theta~\mbox{Tr} W_{A}^{\alpha} W_{A\alpha}+
\int d^2\bar{\theta}~\mbox{Tr} W_{A}^{\dagger\dot{\alpha}} W^{\dagger}_{A\dot{\alpha}}\right) \nonumber \\
+\sum_{i} \int d^2\theta  d^2\bar{\theta} ~\Phi_i^{\dagger} e^{2 V_3 + 2 V_2 + 2 V_1} \Phi_i +
\biggl(\int d^2\theta ~ W_{MSSM}(\Phi_k)  + h.c.\biggr)\,,
\label{mssm9}
\end{eqnarray}
where $\Phi_i$ is a set of chiral superfields that MSSM contains.

In general, the set of the soft SUSY breaking terms includes \cite{Girardello:1981wz} 
\begin{equation}
-{\cal L}_{soft} = \biggl(\frac{1}{2}\sum_{A} M_{A}\tilde{\lambda}_{A}\tilde{\lambda}_{A} +h.c.\biggr)+V_{soft}\,,
\label{mssm10}
\end{equation}
$$
V_{soft} = \sum_{i,j} m^2_{ij} \phi^{\dagger}_i\phi_j 
+\sum_{j,j,k} \biggl(\frac{1}{6} a_{ijk} \phi_i \phi_j \phi_k + \frac{1}{2} b_{ij} \phi_i \phi_j+ t_i \phi_i + 
h.c.\biggr)\,,
$$
where $\phi_i$ are scalar components of chiral superfields $\Phi_i$ and $\tilde{\lambda}_{A}$ are
gauginos of the gauge group associated with index $A$. The terms in ${\cal L}_{soft}$ clearly break 
SUSY because they involve only scalars and gauginos and not their respective superpartners.
A set of soft SUSY breaking parameters includes gaugino masses $M_A$, soft scalar masses $m_{ij}^2$, 
tadpole couplings $t_i$, trilinear and bilinear scalar couplings ($a_{ijk}$ and $b_{ij}$).
The soft terms in ${\cal L}_{soft}$ are capable of giving masses to all of the scalars and gauginos,
even if the gauge bosons and fermions in the corresponding supermultiplets are massless or relatively light.
It is worth to point out that soft SUSY breaking terms do not change SUSY relationships between 
the dimensionless couplings. As a consequence softly broken SUSY theory is free of quadratic 
divergences.

The inclusion of the soft SUSY breaking terms in the MSSM introduces many new parameters that
were not present in the SM. A careful count reveals that there are 105 masses, phases and
mixing angles in the MSSM Lagrangian that cannot be rotated away and that have no counterpart 
in the MSSM. On the other hand most of the new parameters lead to flavor mixing and/or CP violating
effects which are severely constrained by different experiments. In order to avoid potentially 
dangerous processes one can assume that all soft SUSY breaking parameters are real,
trilinear scalar couplings are proportional to the corresponding Yukawa couplings and 
$m^2_{ij}\simeq m^2_i \delta_{ij}$. Applying this receipt we get
\begin{equation}
\begin{array}{lcr}
-{\cal L}^{MSSM}_{soft}& = &\sum_{i} m^2_i |\phi_i|^2+\biggl(\frac{1}{2}
\sum_{A}^{} M_{A}\tilde{\lambda}_{A}\tilde{\lambda}_{A}
+ \sum_{a, b} [A^U_{ab} y^U_{ab}\tilde{Q}_a\tilde{u}^c_b H_2\\[2mm]
&+& A^D_{ab} y^D_{ab}\tilde{Q}_a\tilde{d}^c_b H_1+
A^L_{ab} y^L_{ab} \tilde{L}_a\tilde{e}^c_b H_1]
+ B\mu H_1 H_2 +h.c.\biggr)\,,
\end{array}
\label{mssm11}
\end{equation}
where $\tilde{\lambda}_{3}$, $\tilde{\lambda}_{2}$ and $\tilde{\lambda}_{1}$ 
are gluinos, winos and bino respectively, while $\tilde{Q}_a$, $\tilde{u}^c_a$
$\tilde{d}^c_a$, $\tilde{L}_a$ and $\tilde{e}^c_a$ are scalar components of
the corresponding chiral superfields. To avoid fine--tuning the SUSY breaking 
mass parameters are expected to be in the TeV range.

\begin{figure}
\centering
\includegraphics[width=.6\textwidth]{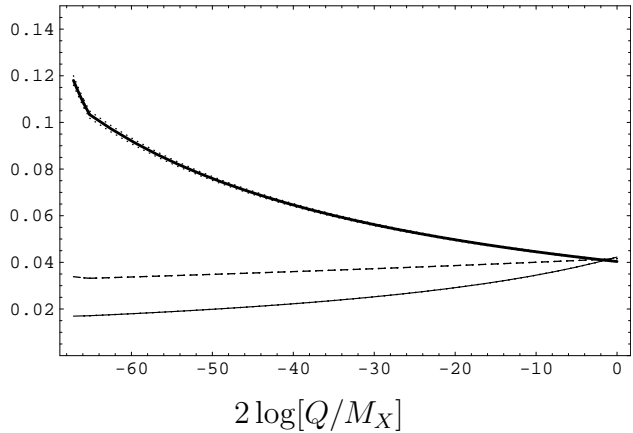}\\
{$2\log[Q/M_X]$}
\caption{Two--loop RG flow of gauge couplings from $M_Z$ to GUT scale $M_X$ 
in the MSSM for $T_S=250\,\mbox{GeV}$, $\alpha_s(M_Z)=0.118$, $\alpha(M_Z)=1/127.9$ and
$\sin^2\theta_W=0.231$. Thick, dashed and solid lines correspond to the running of $SU(3)_C$, 
$SU(2)_W$ and $U(1)_Y$ couplings respectively. The dotted lines represent the uncertainty 
in $\alpha_i(Q)$ caused by the variation of the strong gauge coupling from 0.116 to 0.120 
at the EW scale.}
\label{mssmfig1}
\end{figure}

One of the most appealing features of simplest SUSY extensions of the SM
is the unification of gauge couplings. More than fifteen years ago it was
found that the EW and strong gauge couplings extracted from LEP data
and extrapolated to high energies using the renormalisation group (RG) equation
evolution do not meet within the SM but converge to a common value
at some high energy scale $M_X$ after the inclusion of supersymmetry, e.g. in the
framework of the minimal SUSY standard model \cite{Ellis:1990wk}--\cite{Anselmo:1991uu}. 
This allows one to embed SUSY extensions of the SM into Grand Unified Theories 
(GUTs) \cite{Georgi:1974sy} and superstring ones \cite{Green:1987sp}. Simultaneously, 
the incorporation of weak and strong gauge interactions within GUTs permits to explain the 
peculiar assignment of $U(1)_Y$ charges postulated in the SM and to address the
observed mass hierarchy of quarks and leptons. 

The evolution of gauge couplings within the MSSM is shown in Fig.~1.
Recent studies revealed that the exact unification of gauge couplings in the
minimal SUSY model requires either relatively high effective SUSY threshold 
scale around $1\,\mbox{TeV}$ (which corresponds to $\mu\simeq 6\,\mbox{TeV}$) 
or large values of the $SU(3)_C$ gauge coupling constant $\alpha_3(M_Z)=0.123-0.124$ 
\cite{Langacker:1995fk}--\cite{deBoer:2003xm}. However non--negligible high energy GUT/string 
threshold corrections and/or higher dimension operator effects may facilitate the unification 
of gauge interactions. Due to the lack of direct evidence verifying or falsifying the 
presence of superparticles at low energies, gauge coupling unification remains one 
of the main motivations for low--energy supersymmetry based on experimental data.

\section{Breakdown of gauge symmetry in the simplest SUSY extensions of the SM}

\subsection{EW symmetry breaking in the MSSM}
Including soft SUSY breaking terms and radiative corrections,
the resulting Higgs effective potential in the MSSM can be written as
\begin{eqnarray}
V=m_1^2|H_1|^2+m_2^2|H_2|^2-m_3^2(H_1 H_2+h.c.)+ \frac{g_2^2}{8}\left(H_1^+\sigma_a H_1+H_2^+\sigma_a
H_2\right)^2\nonumber\\
+\frac{{g'}^2}{8}\left(|H_1|^2-|H_2|^2\right)^2+\Delta V\,,\qquad\qquad\qquad\qquad\qquad
\label{7}
\end{eqnarray}
where $g'=\sqrt{3/5} g_1$, $g_2$ and $g_1$ are the low energy (GUT normalised)
$SU(2)_W$ and $U(1)_Y$ gauge couplings, $m_1^2=m_{H_1}^2+\mu^2$, $m_2^2=m_{H_2}^2+\mu^2$
and $m_3^2=-B\mu$. In Eq.~(\ref{7}) $\Delta V$ represents the contribution of loop
corrections to the Higgs effective potential.

At the physical minimum of the scalar potential (\ref{7}) the Higgs
fields develop vacuum expectation values (VEVs)
\begin{equation}
<H_1>=\frac{1}{\sqrt{2}}\left(
\begin{array}{c}
v_1\\ 0
\end{array}
\right) , \qquad
<H_2>=\frac{1}{\sqrt{2}}\left(
\begin{array}{c}
0\\ v_2
\end{array}
\right)\,
\label{8}
\end{equation}
breaking the $SU(2)_W\times U(1)_Y$ gauge symmetry to $U(1)_{em}$ associated
with electromagnetism and generating the masses of all bosons and fermions.
The value of $v=\sqrt{v_1^2+v_2^2}\simeq 246\,\mbox{GeV}$ is fixed
by the Fermi scale. At the same time the ratio of the Higgs VEVs remains
arbitrary. Hence it is convenient to introduce $\tan\beta=v_2/v_1$.

The vacuum configuration (\ref{8}) is not the most general one. Because of
the $SU(2)$ invariance of the Higgs potential (\ref{7}) one can always make
$<H_2^{+}>=0$ by virtue of a suitable gauge rotation. Then the requirement
$<H_1^{-}>=0$, which is a necessary condition to preserve $U(1)_{em}$ 
in the physical vacuum, is equivalent to requiring the squared mass of the 
physical charged scalar to be positive. It imposes additional constraints 
on the parameter space.

At tree--level $\Delta V=0$ and the Higgs potential is described by the sum of
the first five terms in Eq.~(\ref{7}). Notice that the Higgs self--interaction
couplings in Eq.~(\ref{7}) are fixed and determined by the gauge couplings in
contrast with the SM. At the tree--level the MSSM Higgs potential contains
only three independent parameters: $m_1^2$,\,$m_2^2$,\,$m_3^2$.
The stable vacuum of the scalar potential (\ref{7}) exists only if
\begin{equation}
m_1^2+m_2^2> 2|m_3|^2\,.
\label{9}
\end{equation}
This can be easily understood if one notice that in the limit $v_1^2=v_2^2$ the
quartic terms in the Higgs potential vanish. In the considered case the scalar
potential (\ref{7}) remains positive definite only if the condition (\ref{9})
is satisfied. Otherwise physical vacuum becomes unstable, i.e. $v_1^2=v_2^2$ 
tends to be much larger than the EW scale. On the other hand
Higgs doublets acquire non-zero VEVs only when
\begin{equation}
m_1^2 m_2^2 < |m_3|^4\,.
\label{10}
\end{equation}
Indeed, if $m_1^2 m_2^2 > |m_3|^4$ then all Higgs fields have positive masses
for $v_1=v_2=0$ and the breakdown of EW symmetry does not take place.
The conditions (\ref{9}) and (\ref{10}) also follow from the equations for the
extrema of the Higgs boson potential. At tree--level the minimization conditions
in the directions (\ref{8}) in field space read:
\begin{eqnarray}
\frac{\partial V}{\partial v_1}=\left(m_1^2+\frac{\bar{g}^2}{8}(v_1^2-v_2^2)\right)v_1-m_3^2 v_2=0\,,\nonumber\\
\frac{\partial V}{\partial v_2}=\left(m_2^2+\frac{\bar{g}^2}{8}(v_2^2-v_1^2)\right)v_2-m_3^2 v_1=0\,,
\label{11}
\end{eqnarray}
where $\bar{g}=\sqrt{g_2^2+g'^2}$. Solution of the minimization conditions
(\ref{11}) can be written in the following form
\begin{eqnarray}
\sin 2\beta=\frac{2 m_3^2}{m_1^2+m_2^2}\,,\qquad\qquad
\frac{\bar{g}^2}{4}v^2=\frac{2(m_1^2-m_2^2\tan^2\beta)}{\tan^2\beta-1}\,.
\label{12}
\end{eqnarray}
Requiring that $v^2>0$ and $|\sin 2\beta|<1$ one can reproduce conditions (\ref{9})
and (\ref{10}).

From Eqs.~(\ref{9})--(\ref{12}) it is easy to see that the appropriate breakdown of
the EW symmetry can not be always achieved. In particular the breakdown of
$SU(2)_W\times U(1)_Y \to U(1)_{em}$ does not take place when $m_2^2=m_1^2$.
This is exactly what happens in the constrained version of the MSSM (cMSSM) which
implies that $m_i^2(M_X)=m_0^2$,\,\,$A^k_{ab}(M_X)=A$,\,\,$M_{\alpha}(M_X)=M_{1/2}$.
Thus in the cMSSM $m_2^2(M_X)=m_1^2(M_X)$. Nevertheless the correct pattern
of the EW symmetry breaking can be achieved within this SUSY model.
Since the top--quark Yukawa coupling is large it affects the renormalisation
group flow of $m_2^2(Q)$ rather strongly resulting in small or even negative
values of $m_2^2(Q)$ at low energies that triggers the breakdown of the
EW symmetry. This is the so-called radiative mechanism of the EW symmetry
breaking (EWSB) \cite{Ibanez:1982fr}--\cite{AlvarezGaume:1983gj}.

The radiative EWSB demonstrates the importance of loop effects in the considered
process. In addition to the RG flow of all couplings one has to take into account
loop corrections to the Higgs effective potential which are associated with the
last term $\Delta V$ in Eq.~(\ref{7}). In the simplest SUSY extensions of the SM
the dominant contribution to $\Delta V$ comes from the loops involving the top--quark
and its superpartners because of their large Yukawa coupling $h_t$. Since in SUSY
theories each fermion state with a specific chirality has a superpartner $t$--quark
has two scalar superparners ($\tilde{t}_L$ and $\tilde{t}_R$) associated with the
left--handed and right--handed top quark states. Due to the EWSB $\tilde{t}_L$ and
$\tilde{t}_R$ get mixed resulting in the formation of two charged scalar particles
with masses
\begin{eqnarray}
m_{\tilde{t}_{1,2}}=\frac{1}{2}\left(m^2_Q+m^2_U+2m_t^2 \pm
\sqrt{(m_Q^2-m_U^2)^2+4m_t^2 \, X_t^2}\right)\,,
\label{13}
\end{eqnarray}
where $X_t=A_t-\mu/\tan\beta$ is a stop mixing parameter, $A_t$ is a trilinear
scalar coupling associated with the top quark Yukawa coupling and $m_t$ is the
running top quark mass
$$
m_t(M_t)=\frac{h_t(M_t)}{\sqrt{2}}v\sin\beta\,.
$$
In the one--loop approximation the contribution of the top--quark
and its superpartners to $\Delta V$ is determined by the masses of the
corresponding bosonic and fermionic states, i.e.
\begin{eqnarray}
\Delta V=\frac{3}{32\pi^2}\left[m_{\tilde{t}_1}^4\left(
\ln\frac{m_{\tilde{t}_1}^2}{Q^2}-\frac{3}{2}\right)+
m_{\tilde{t}_2}^4\left(\ln\frac{m_{\tilde{t}_2}^2}{Q^2}-\frac{3}{2}\right)\qquad\right.\nonumber\\[2mm]
\left.\qquad-2m_t^4\left(\ln\frac{m_t^2}{Q^2}-\frac{3}{2}\right)\right]\,.
\label{14}
\end{eqnarray}

Initially the sector of EWSB involves eight degrees of freedom. However three
of them are massless Goldstone modes which are swallowed by the $W^{\pm}$ and
$Z$ gauge bosons. The $W^{\pm}$ and $Z$ bosons gain masses via the interaction
with the neutral components of the Higgs doublets so that
$$
M_W=\frac{g_2}{2}v\,,\qquad\qquad M_Z=\frac{\bar{g}}{2}v\,.
$$

When CP in the MSSM Higgs sector is conserved the remaining five
physical degrees of freedom form two charged, one CP--odd and two CP-even
Higgs states. The masses of the charged and CP-odd Higgs bosons are
\begin{equation}
m_{A}^2=m_1^2+m_2^2+\Delta_A\,,\qquad\qquad
M_{H^{\pm}}^2=m_A^2+M_W^2+\Delta_{\pm}\,,
\label{15}
\end{equation}
where $\Delta_{\pm}$ and $\Delta_A$ are the loop corrections.
The CP--even states are mixed and form a $2\times 2$ mass matrix.
It is convenient to introduce a new field space basis $(h,\,H)$
rotated by the angle $\beta$ with respect to the initial one:
\begin{eqnarray}
\mbox{Re} \, H_1^0= (h \cos\beta- H \sin\beta+v_1)/\sqrt{2}\,,\nonumber\\
\mbox{Re} \, H_2^0= (h \sin\beta+ H \cos\beta+v_2)/\sqrt{2}\,.
\label{16}
\end{eqnarray}
In this new basis the mass matrix of the Higgs scalars takes the form
\cite{Kovalenko:1998dc}
\begin{equation}
M^2=\left(
\begin{array}{ll}
M^2_{11}\quad & M^2_{12}\\[3mm]
M^2_{21}\quad & M^2_{22}
\end{array}
\right)=
\left(
\begin{array}{ll}
\displaystyle\frac{\partial^2 V}{\partial v^2} &
\displaystyle\frac{1}{v}\frac{\partial^2V}{\partial v \partial\beta}\\[3mm]
\displaystyle\frac{1}{v}\frac{\partial^2V}{\partial v \partial\beta} &
\displaystyle\frac{1}{v^2}\frac{\partial^2V}{\partial\beta^2}
\end{array}
\right)\,,
\label{17}
\end{equation}
\begin{eqnarray}
M_{22}^2=m_A^2+M_Z^2\sin^2 2\beta+\Delta_{22}\, ,\qquad 
M_{11}^2=M_Z^2\cos^2 2\beta+\Delta_{11}\,,
\label{18}
\end{eqnarray}
$$
M_{12}^2=M_{21}^2=-\frac{1}{2} M_Z^2\sin 4\beta+\Delta_{12}\,,
$$
where $\Delta_{ij}$ represent the contributions from loop corrections.
In Eqs.~(\ref{18}) the equations for the extrema of the Higgs
boson effective potential are used to eliminate $m_1^2$ and $m_2^2$.

From Eqs.~(\ref{15}) and (\ref{18}) one can see that at tree--level
the masses and couplings of the Higgs bosons in the MSSM can be
parametrised in terms $m_A$ and $\tan\beta$ only. The masses of the
two CP--even eigenstates obtained by diagonalizing the matrix
(\ref{17})--(\ref{18}) are given by
\begin{equation}
m_{h_1,\,h_2}^2=\frac{1}{2}\left(M^2_{11}+
M^2_{22}\mp\sqrt{(M_{22}^2-M_{11}^2)^2+4M^4_{12}}\right)\,.
\label{19}
\end{equation}
The qualitative pattern of the Higgs spectrum depends very strongly on
the mass $m_A$ of the pseudoscalar Higgs boson. With increasing $m_A$
the masses of all the Higgs particles grow. At very large values of
$m_A$ ($m_A^2\gg v^2$), the lightest Higgs boson mass approaches its
theoretical upper limit $\sqrt{M_{11}^2}$. Thus the top--left entry
of the CP--even mass matrix (\ref{17})--(\ref{18}) represents an upper
bound on the lightest Higgs boson mass--squared. At the tree--level
the lightest Higgs boson mass is always less than the $Z$--boson mass:
$m_{h_1}\lesssim M_Z^2\cos^2 2\beta$ \cite{Inoue:1982ej}--\cite{Flores:1982pr}.
In the leading two--loop approximation the mass of the lightest Higgs boson 
in the MSSM does not exceed $130-135\,\mbox{GeV}$. This is one of the most 
important predictions of the minimal SUSY model that can be tested at the LHC
in the near future.

It is important to study the spectrum of the Higgs bosons together
with their couplings to the gauge bosons because LEP sets stringent
limits on the masses and couplings of the Higgs states. In the rotated
field basis $(h, H)$ the trilinear part of the Lagrangian, which determines
the interactions of the neutral Higgs states with the $Z$--boson,
is simplified:
\begin{equation}
L_{AZH}=\frac{\bar{g}}{2}
M_{Z}Z_{\mu}Z_{1\mu}h+\frac{\bar{g}}{2}Z_{\mu}
\biggl[H(\partial_{\mu}A)-(\partial_{\mu}H)A\biggr]~.
\label{20}
\end{equation}
In this basis the SM-like CP--even component $h$ couples to a pair of
$Z$ bosons, while the other one $H$ interacts with the pseudoscalar $A$
and $Z$. The coupling of $h$ to the $Z$ pair is exactly the same as in
the SM. The couplings of the Higgs mass eigenstates to a $Z$ pair
($g_{ZZh_i}$, $i=1,2$) and to the Higgs pseudoscalar and $Z$ boson
($g_{ZAh_i}$) appear because of the mixing between $h$ and $H$.

Following the traditional notations, one can define the normalised
$R$--couplings as: $g_{VVh_i}=R_{VVh_i}\times \mbox{SM coupling}$
$\left(V=Z,\,W^{\pm}\right)$; $g_{ZAh_i}=\frac{\bar{g}}{2}R_{ZAh_i}$.
The absolute values of all these $R$--couplings vary from zero to unity.
The relative couplings $R_{ZZh_i}$ and $R_{ZAh_i}$ are given in terms
of the angles $\alpha$ and $\beta$:
\begin{equation}
R_{VVh_1}=-R_{ZAh_2}=\sin(\beta-\alpha)\,,\quad R_{VVh_2}=R_{ZAh_1}=\cos(\beta-\alpha)\,,
\label{21}
\end{equation}
where the angle $\alpha$ is defined as follows:
\begin{equation}
\begin{array}{rcl}
h_1&=&-(\sqrt{2}\, \mbox{Re}\,H_1^0-v_1)\sin\alpha+(\sqrt{2}\, \mbox{Re}\,H_2^0-v_2)\cos\alpha\,,\\[1mm]
h_2&=&(\sqrt{2}\, \mbox{Re}\,H_1^0-v_1)\cos\alpha+(\sqrt{2}\, \mbox{Re}\,H_2^0-v_2)\sin\alpha\,.
\end{array}
\label{22}
\end{equation}
From Eqs.~(\ref{21}) it becomes clear that in the MSSM the couplings of the
lightest Higgs boson to the Z pair can be substantially smaller than in the SM.
Therefore the experimental lower bound on the lightest Higgs mass in the MSSM is
weaker than in the SM. On the other hand, when $m_A^2\gg v^2$ the lightest
CP--even Higgs state is predominantly the SM-like superposition of the neutral
components of Higgs doublets $h$ so that $R_{ZZh_1}\simeq 1$. In this case
the lightest Higgs scalar has to satisfy LEP constraint on the mass of the SM--like
Higgs boson, i.e. it should be heavier than $114.4\,\mbox{GeV}$.

\begin{figure}
\centering
~\hspace*{-6.1cm}{$m_{h_1}$}\hspace{6.8cm}{$m_{h_1}$}\\
\includegraphics[width=0.43\textwidth]{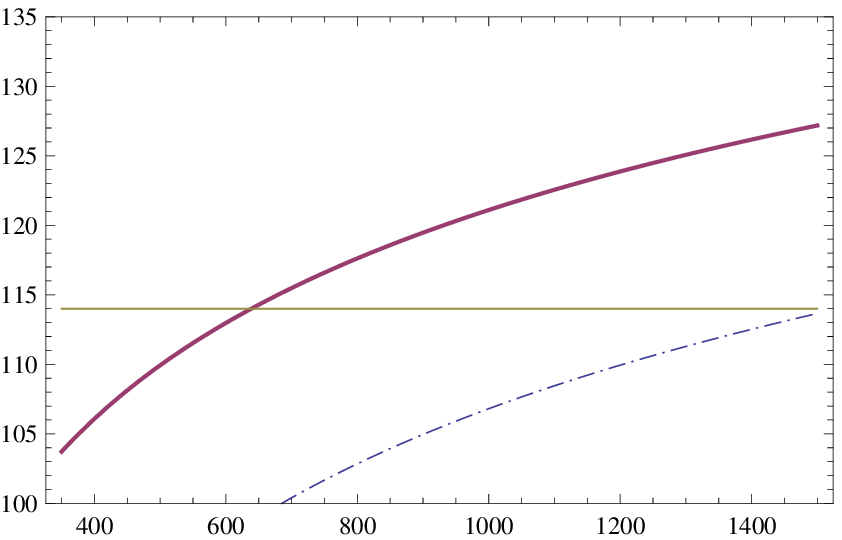}\qquad
\includegraphics[width=0.43\textwidth]{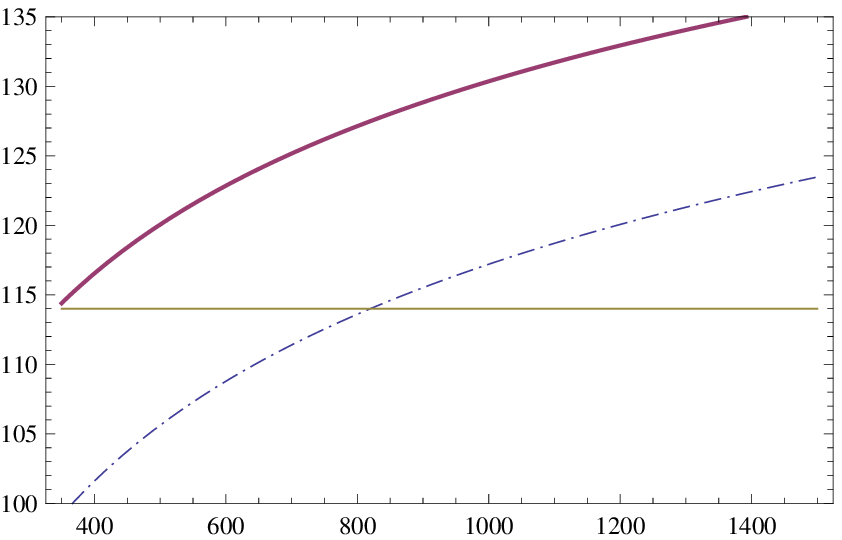}\\
~\hspace*{-0.1cm} {$M_S$}\hspace{6.8cm}{$M_S$}
\caption{{\it Left:} The dependence of the one--loop lightest Higgs boson mass
on $M_S$ for $\tan\beta=2$. {\it Right:} The one--loop mass of the lightest
CP--even Higgs state versus $M_S$ for $\tan\beta=3$. The solid and dashed--dotted
lines correspond to $X_t=2\,M_S$ and $X_t= M_S$ respectively. The horizontal line
represents the current LEP limit.}
\label{figure-1}
\end{figure}

Recent studies indicate that in the MSSM the scenarios with the light Higgs
pseudoscalar ($m_A\sim 100\,\mbox{GeV}$) are almost ruled out by LEP. Since
$m_A$ tend to be large the SM--like Higgs boson must be relatively heavy.
At the same time, as it has been already mentioned, the tree--level mass of
the lightest Higgs boson in the MSSM does not exceed $M_Z\simeq 91\,\mbox{GeV}$. 
As a consequence in order to satisfy LEP constraints large contribution of 
loop corrections to the mass of the lightest CP--even Higgs state is required. 
When SUSY breaking scale $M_S$ is considerably larger than $M_Z$ and 
$m_Q^2\simeq m_U^2\simeq M_S^2$ the contribution of the one--loop corrections 
to $m_{h_1}^2$ in the leading approximation can be written as
\begin{equation}
\Delta^{(1)}_{11}\simeq\frac{3 M_t^4 }{2\pi^2 v^2}
\left[\frac{X_t^2}{M_S^2}\biggl(1-\frac{1}{12}\frac{X_t^2}{M_S^2}\biggr)+
\ln\biggl(\frac{M_S^2}{m_t^2}\biggr)\right]\,.
\label{23}
\end{equation}
The large values of $\Delta^{(1)}_{11}\sim M_{Z}^2$ can be obtained only
if $M_S\gg m_t$ and the ratio $|X_t/M_S|$ is also large. The contribution of
the one--loop corrections (\ref{23}) attains its maximal value for $X_t^2=6\,M_S^2$.
This is the so--called maximal mixing scenario. In Figs.~2a and 2b the dependence of
the one--loop lightest Higgs boson mass on the SUSY breaking scale $M_S$ for $\tan\beta=2$
and $\tan\beta=3$ is examined. Two different cases $X_t=M_S$ and $X_t=2\,M_S$ are
considered. From Figs.~2a and 2b one can see that in order to satisfy LEP constraints
$M_S$ should be larger than $400-600\,\mbox{GeV}$. Leading two--loop corrections
reduce the SM--like Higgs mass even further. As a result larger values of the
SUSY breaking scale are required to overcome LEP limit.

Large values of the soft scalar masses of the superpartners of the top quark
($m_Q^2,\,m_U^2 \gg M_Z^2$) tend to induce large and negative value of $m_{H_2}^2$ 
in the Higgs potential due to the RG flow. This leads to the fine tuning because 
$m_1^2$ and $m_2^2$ determine the EW scale (see Eqs.~(\ref{12})).
Generically the fine tuning which is required to overcome LEP constraints
in the MSSM is of the order of 1\% (little hierarchy problem). This fine
tuning should be compared with the fine tuning in other theories which are
used in particle physics. In particular, it is well known that QCD is
a highly fine--tuned theory. Indeed, QCD Lagrangian should contain ``$\theta$-term''
\begin{equation}
{\cal L}_{\theta} = \theta_{\rm eff}
\frac{\alpha_s}{8 \pi} F^{\mu \nu \, a} \tilde{F}_{\mu \nu}^a,
\qquad\qquad \theta_{\rm eff} = \theta + {\rm arg \; det}\; M_q\,,
\label{24}
\end{equation}
where $F^{\mu \nu \, a}$ is the gluon field strength and
$\tilde{F}_{\mu \nu}^a \equiv\frac{1}{2}\epsilon_{\mu\nu\rho\sigma}F^{\rho\sigma\, a}$
is its dual. This term is not forbidden by the gauge invariance. On the other
hand the parameter $\theta$ must be extremely small, i.e.
$|\theta_{\rm eff}| \lesssim 10^{-9}$ (strong CP problem). Otherwise it results
in too large value of the neutron electric dipole moment. Eqs.~(\ref{24})
demonstrate that so small value of $\theta_{\rm eff}$ implies enormous
fine tuning which is much higher than in the MSSM.

The little hierarchy problem can be solved within SUSY models that allow to get
relatively large mass of the SM--like Higgs boson ($m_{h_1}\gtrsim 100-110\,\mbox{GeV}$)
at the tree-level. Alternatively, one can try to avoid stringent LEP constraints
by allowing exotic decays of the lightest Higgs particle. If usual branching ratios
of the lightest Higgs state are dramatically reduced then the lower LEP bound
on the SM--like Higgs mass may become inapplicable. In this case the lightest
Higgs boson can be still relatively light so that large contribution of
loop corrections is not required. Both possibilities mentioned above imply
the presence of new particles and interactions. These new particles and
interactions can be also used to solve the so-called $\mu$ problem.
This problem arises when MSSM gets incorporated into supergravity and/or
GUT models. Within these models the parameter $\mu$ is expected to be
either zero or of the order of Planck/GUT scale. At the same time in order
to provide the correct pattern of the EWSB $\mu$ is required to be of the
order of the EW scale.

\subsection{Higgs sector of the NMSSM}
In the simplest extension of the MSSM, the Next--to--Minimal Supersymmetric Standard
Model (NMSSM), the superpotential is invariant with respect to the discrete
transformations $\Phi_{i}\to e^{2\pi i/3}\Phi_{i}$
of the $Z_3$ group (for recent review see \cite{Ellwanger:2009dp}). The term $\mu (H_1 H_2)$
does not meet this requirement. Therefore it is replaced in the superpotential by
\begin{equation}
W_{H}=\lambda S (H_1 H_2)+\frac{1}{3}\kappa S^3\,,
\label{25}
\end{equation}
where $S$ is an additional superfield which is a singlet with respect to $SU(2)_W$
and $U(1)_Y$ gauge transformations. A spontaneous breakdown of the EW symmetry
leads to the emergence of the VEV of extra singlet field $<S>=s/\sqrt{2}$ and an
effective $\mu$ parameter is generated ($\mu=\lambda s/\sqrt{2}$).

The potential energy of the Higgs field interaction can be written as a sum
\begin{equation}
V=V_F+V_D+V_{soft}+\Delta V\, ,
\label{26}
\end{equation}
\begin{eqnarray}
V_F=\lambda^2|S|^2(|H_1|^2+|H_2|^2)+\lambda^2|(H_1 H_2)|^2\qquad\qquad \nonumber\\[2mm]
\qquad\qquad+\lambda\kappa\left[S^{*2}(H_1 H_2)+h.c.\right]+\kappa^2|S|^4\, ,
\label{27}
\end{eqnarray}
\begin{equation}
V_D=\frac{g_2^2}{8}\left(H_1^+\sigma_a H_1+H_2^+\sigma_a
H_2\right)^2+\frac{{g'}^2}{8}\left(|H_1|^2-|H_2|^2\right)^2\, ,
\label{28}
\end{equation}
\begin{eqnarray}
V_{soft}=m_1^2|H_1|^2+m_2^2|H_2|^2+m_S^2|S|^2\qquad\qquad \nonumber\\[2mm]
\qquad\qquad +\left[\lambda A_{\lambda}S(H_1 H_2)+\frac{\kappa}{3}A_{\kappa}S^3+h.c.\right]\, .
\label{29}
\end{eqnarray}
At the tree level the Higgs potential (\ref{26}) is described by the sum of the first three terms.
$V_F$ and $V_D$ are the $F$ and $D$ terms. Their structure is fixed by the superpotential (\ref{25})
and the EW gauge interactions in the common manner. The soft SUSY breaking terms are collected in
$V_{soft}$. The set of soft SUSY breaking parameters involves soft masses $m_1^2,\, m_2^2,\, m_S^2$
and trilinear couplings $A_{\kappa},\, A_{\lambda}$. The last term in Eq.~(\ref{26}), $\Delta V$,
corresponds to the contribution of loop corrections. In the leading one--loop approximation
$\Delta V$ in the NMSSM is given by Eqs.~(\ref{13})--(\ref{14}) in which $\mu$ has to be replaced
by $\lambda s/\sqrt{2}$. Further we assume that $\lambda$, $\kappa$ and all soft SUSY breaking
parameters are real so that CP is conserved.

At the physical vacuum of the Higgs potential
\begin{equation}
<H_1>=\frac{1}{\sqrt{2}}\left(
\begin{array}{c}
v_1\\ 0
\end{array}
\right) , \quad
<H_2>=\frac{1}{\sqrt{2}}\left(
\begin{array}{c}
0\\ v_2
\end{array}
\right)\,,\quad <S>=\frac{s}{\sqrt{2}}\,.
\label{30}
\end{equation}
The equations for the extrema of the full Higgs boson effective potential in the directions 
(\ref{30}) in the field space are given by
\begin{eqnarray}
\frac{\partial V}{\partial s}=\left(m_S^2+\frac{\lambda^2}{2}(v_1^2+v_2^2)-\lambda\kappa v_1 v_2
\right)s-\frac{\lambda A_{\lambda}}{\sqrt{2}}v_1v_2 \qquad\qquad\nonumber\\[2mm]
\qquad\qquad+\frac{\kappa A_{\kappa}}{\sqrt{2}}s^2+
\kappa^2 s^3+\frac{\partial \Delta V}{\partial s}=0\, ,
\label{31}
\end{eqnarray}
\begin{eqnarray}
\frac{\partial V}{\partial v_1}=\left(m_1^2+\frac{\lambda^2}{2}(v_2^2+s^2)+
\frac{\bar{g}^2}{8}(v_1^2-v_2^2)\right)v_1\qquad\qquad\qquad\nonumber\\[2mm]
\qquad\qquad-\left(\frac{\lambda\kappa}{2}s^2+\frac{\lambda A_{\lambda}}{\sqrt{2}}s\right)v_2
+\frac{\partial \Delta V}{\partial v_1}=0\, ,
\label{32}
\end{eqnarray} 
\begin{eqnarray}
\frac{\partial V}{\partial v_2}=\left(m_2^2+\frac{\lambda^2}{2}
(v_1^2+s^2)+\frac{\bar{g}^2}{8}(v_2^2-v_1^2)\right)v_2\qquad\qquad\qquad\nonumber\\[2mm]
\qquad\qquad-\left(\frac{\lambda\kappa}{2}s^2+\frac{\lambda A_{\lambda}}{\sqrt{2}}s\right)v_1
+\frac{\partial \Delta V}{\partial v_2}=0\, .
\label{33}
\end{eqnarray}
As in the MSSM upon the breakdown of the EW symmetry three goldstone modes ($G^{\pm}$ and $G^{0}$)
emerge, and are absorbed by the $W^{\pm}$ and $Z$ bosons. In the field space basis rotated by an
angle $\beta$ with respect to the initial direction, i.e.
\begin{eqnarray}
\mbox{Im}\, H_1^0 = (P \sin \beta + G^0 \cos \beta)/\sqrt{2}\, ,\qquad & 
H_2^+=H^+ \cos \beta - G^+ \sin \beta\,, \nonumber\\[0mm]
\mbox{Im}\, H_2^0 = (P \cos \beta - G^0 \sin \beta)/\sqrt{2} \, ,\qquad& 
H_1^-=G^- \cos \beta + H^- \sin \beta\,, \nonumber\\[0mm]
\mbox{Re} \, H_1^0= (h \cos\beta- H \sin\beta+v_1)/\sqrt{2}\,,\, & 
\mbox{Im}\, S = P_S/\sqrt{2}\,, \qquad\qquad\qquad\nonumber\\[0mm]
\mbox{Re} \, H_2^0= (h \sin\beta+ H \cos\beta+v_2)/\sqrt{2}\,,\, & 
\mbox{Re} \, S = (s+N)/\sqrt{2}\, ,\qquad\qquad
\label{34}
\end{eqnarray}
these unphysical degrees of freedom decouple and the mass terms in the Higgs boson potential
can be written as follows
\begin{equation}
V_{mass} = M_{H^{\pm}}^2  H^+ H^- +
\frac{1}{2} (P \,\, P_S)\, \tilde{M}^2
\left(
\begin{array}{c}
P \\
P_S
\end{array}
\right)+
\frac{1}{2} (H \,\, h \,\, N)\, M^2
\left(
\begin{array}{c}
H \\
h \\
N
\end{array} \right)\, .
\label{35}
\end{equation}

From the conditions for the extrema (\ref{31})--(\ref{33}) one can express $m_S^2$, $m_1^2$,
$m_2^2$ via other fundamental parameters, $\tan\beta$ and $s$. Substituting the obtained
relations for the soft masses in the $2\times 2$ CP-odd mass matrix $\tilde{M}^2_{ij}$ we get:
$$
\tilde{M}_{11}^2=m_A^2=\frac{4\mu^2}{\sin^2 2\beta}\left(x-\frac{\kappa}{2\lambda}\sin2\beta\right)
+\tilde{\Delta}_{11}\, ,
$$
\begin{equation}
\tilde{M}_{22}^2=\frac{\lambda^2 v^2}{2}x+\frac{\lambda\kappa}{2}v^2\sin2\beta-
3\frac{\kappa}{\lambda}A_{\kappa}\mu+\tilde{\Delta}_{22}\, ,
\label{36}
\end{equation}
$$
\tilde{M}_{12}^2=\tilde{M}_{21}^2=\sqrt{2}\lambda v \mu\left(\frac{x}{\sin 2\beta}
-2\frac{\kappa}{\lambda}\right)+\tilde{\Delta}_{12}\, ,
$$
where $x=\displaystyle\frac{1}{2\mu}\left(A_{\lambda}+2\frac{\kappa}{\lambda}\mu\right)\sin2\beta$,
$\mu=\displaystyle\frac{\lambda s}{\sqrt{2}}$ and $\tilde{\Delta}_{ij}$
are contributions of the loop corrections to the mass matrix elements.
The mass matrix (\ref{36}) can be easily diagonalized.
The corresponding eigenvalues are given by
\begin{equation}
m^2_{A_2, A_1}=\frac{1}{2}\left(\tilde{M}^2_{11}+\tilde{M}^2_{22}\pm
\sqrt{(\tilde{M}^2_{11}-\tilde{M}^2_{22})^2+4\tilde{M}^4_{12}}
\right)~.
\label{38}
\end{equation}

Because the charged components of the Higgs doublets are not mixed with the neutral Higgs
states the charged Higgs fields $H^{\pm}$ are already physical mass eigenstates with
\begin{equation}
M_{H^{\pm}}^2=m_A^2-\frac{\lambda^2 v^2}{2}+M_W^2+\Delta_{\pm}.
\label{39}
\end{equation}
Here $\Delta_{\pm}$ includes loop corrections to the charged Higgs mass.

In the rotated basis  $H\, ,h\,, N$ the mass matrix of the CP--even Higgs sector
takes the form \cite{Kovalenko:1998dc},\cite{Nevzorov:2001um}:
\begin{equation}
M^2=\left(
\begin{array}{ccc}
M_{11}^2 & M_{12}^2 & M_{13}^2\\[2mm]
M_{21}^2 & M_{22}^2 & M_{23}^2\\[2mm]
M_{31}^2 & M_{32}^2 & M_{33}^2
\end{array}
\right)=
\left(
\begin{array}{ccc}
\displaystyle\frac{1}{v^2}\frac{\partial^2 V}{\partial^2\beta}&
\displaystyle\frac{1}{v}\frac{\partial^2 V}{\partial v \partial\beta}&
\displaystyle\frac{1}{v}\frac{\partial^2 V}{\partial s \partial\beta}\\[2mm]
\displaystyle\frac{1}{v}\frac{\partial^2 V}{\partial v \partial\beta}&
\displaystyle\frac{\partial^2 V}{\partial v^2}&
\displaystyle\frac{\partial^2 V}{\partial v \partial s}\\[2mm]
\displaystyle\frac{1}{v}\frac{\partial^2 V}{\partial s \partial\beta}&
\displaystyle\frac{\partial^2 V}{\partial v \partial s}&
\displaystyle\frac{\partial^2 V}{\partial^2 s}
\end{array}
\right)~,
\label{40}
\end{equation}
\begin{equation}
\begin{array}{rcl}
M_{11}^2&=&\displaystyle m_A^2+\left(\frac{\bar{g}^2}{4}-\frac{\lambda^2}{2}\right)v^2
\sin^2 2\beta+\Delta_{11}\, ,\\
M_{22}^2&=&\displaystyle M_Z^2\cos^2 2\beta+\frac{\lambda^2}{2}v^2\sin^2 2\beta+
\Delta_{22}\, ,\\
M_{33}^2&=&\displaystyle 4\frac{\kappa^2}{\lambda^2}\mu^2+\frac{\kappa}{\lambda}A_{\kappa}\mu+
\frac{\lambda^2 v^2}{2}x-\frac{\kappa\lambda}{2}v^2\sin2\beta+\Delta_{33}\, ,\\
M_{12}^2&=&M_{21}^2=\displaystyle \left(\frac{\lambda^2}{4}-\frac{\bar{g}^2}{8}\right)v^2
\sin 4\beta+\Delta_{12}\, ,\\
M_{13}^2&=&M_{31}^2=-\displaystyle \frac{\sqrt{2}\lambda v \mu x \cos 2\beta}{\sin 2\beta} +\Delta_{13}\, ,\\
M_{23}^2&=&M_{32}^2=\sqrt{2}\lambda v \mu (1-x)+\Delta_{23}\, ,
\end{array}
\label{41}
\end{equation}
where $\Delta_{ij}$ can be calculated by differentiating $\Delta V$.

At least one Higgs state in the CP--even sector is always light. Since the minimal eigenvalue of
a Hermitian matrix does not exceed its smallest diagonal element the lightest CP--even Higgs boson squared
mass $m_{h_1}^2$ remains smaller than $M_{22}^2\sim M_Z^2$ even when the supersymmetry breaking
scale is much larger than the EW scale\footnote{The same theorem may lead to the upper bound on the
mass of the lightest neutralino \cite{Hesselbach:2007te}.}. At the tree--level the upper bound on 
the lightest Higgs mass in the NMSSM was found in \cite{Durand:1988rg}--\cite{Drees:1988fc}. 
It differs from the theoretical bound in the MSSM only for moderate values of $\tan\beta$. 
In the leading two--loop approximation the lightest Higgs boson mass in the NMSSM does not exceed 
$135\,\mbox{GeV}$. As follows from the explicit form of the mass matrices (\ref{36}) and (\ref{41}) 
at the tree-level, the spectrum of the Higgs bosons and their couplings depend on the six parameters:
$\lambda, \kappa, \mu, \tan\beta, A_{\kappa}$ and $m_A$ (or $x$). 

First let us consider the MSSM limit of the NMSSM.
Because the strength of the interaction of the extra SM singlet superfield $S$ with $H_1$ and $H_2$
is determined by the size of the coupling $\lambda$ in the superpotential (\ref{25}) the MSSM expressions
for the Higgs masses and couplings are reproduced when $\lambda$ tends to be zero. On the other hand
the equations (\ref{32})--(\ref{33}) imply that $s$ should grow with decreasing $\lambda$ as $M_Z/\lambda$
to ensure the correct breakdown of the EW symmetry. In the limit $\lambda\to 0$ all terms, which are
proportional to $\lambda v_i$, in the minimization conditions (\ref{31}) can be neglected and the
corresponding equation takes the form:
\begin{equation}
s\left(m_S^2+\frac{\kappa A_{\kappa}}{\sqrt{2}}s
+\kappa^2 s^2\right)\simeq 0
\label{44}
\end{equation}
The Eq.~(\ref{44}) has always at least one solution $s_0=0$. In addition two non-trivial roots arise
if $A_{\kappa}^2> 8 m_S^2$. They are given by
\begin{equation}
s_{1,2}\simeq\frac{-A_{\kappa}\pm\sqrt{A_{\kappa}^2-8 m_S^2}}{2\sqrt{2}\kappa}\,\,.
\label{45}
\end{equation}
When $m_S^2>0$ the root $s_0=0$ corresponds to the local minimum of the Higgs potential (\ref{26})--(\ref{29})
that does not lead to the acceptable solution of the $\mu$--problem. The second non-trivial vacuum, that appears if
$A_{\kappa}^2> 8 m_S^2$, remains unstable for $A_{\kappa}^2< 9 m_S^2$. Larger absolute values of $A_{\kappa}$
$(A_{\kappa}^2> 9 m_S^2)$ stabilizes the second minimum which is attained at $s=s_1 (s_2)$ for negative
(positive) $A_{\kappa}$. From Eq.~(\ref{45}) it becomes clear that the increasing of $s$ can be achieved either
by decreasing $\kappa$ or by raising $m_S^2$ and $A_{\kappa}$. Since there is no natural reason why $m_S^2$ and
$A_{\kappa}$ should be very large while all other soft SUSY breaking terms are left in the TeV range, the values
of $\lambda$ and $\kappa$ are obliged to go to zero simultaneously so that their ratio remains unchanged.

Since in the MSSM limit of the NMSSM mixing between singlet states and neutral components of the Higgs doublets
is small the mass matrices (\ref{36}) and (\ref{40})--(\ref{41}) can be diagonalised using the perturbation
theory \cite{Kovalenko:1998dc},\cite{Nevzorov:2001um},\cite{Nevzorov:2000uv}--\cite{Nevzorov:2004ge}.
At the tree--level the masses of two Higgs pseudoscalars are given by
\begin{equation}
m_{A_2}^2\simeq m_A^2=\frac{4\mu^2}{\sin^2 2\beta}\left(x-\frac{\kappa}{2\lambda}\sin2\beta\right)\,, \qquad
m_{A_1}^2\simeq-3\frac{\kappa}{\lambda}A_{\kappa}\mu\,.
\label{46}
\end{equation}
The masses of two CP--even Higgs bosons are the same as in the MSSM (see Eq.~(\ref{19})) while the mass of the extra
CP--even Higgs state, which is predominantly a SM singlet field, is set by $\frac{\kappa}{\lambda}\mu$
\begin{equation}
m_{h_3}^2\approx 4\frac{\kappa^2}{\lambda^2}\mu^2+\frac{\kappa}{\lambda}A_{\kappa}\mu
+\frac{\lambda^2v^2}{2}x\sin^22\beta-\frac{2\lambda^2v^2\mu^2(1-x)^2}{M_Z^2\cos^22\beta}\, .
\label{47}
\end{equation}
The parameter $A_{\kappa}$ occurs in the masses of extra scalar $m_{h_3}$ and pseudoscalar $m_{A_1}$ with opposite
sign and is therefore responsible for their splitting. To ensure that the physical vacuum is a global minimum of the
Higgs potential (\ref{26})--(\ref{29}) and the masses-squared of all Higgs states are positive in this
vacuum the parameter $A_{\kappa}$ must satisfy the following constraints
\begin{equation}
-3\left(\frac{\kappa}{\lambda}\mu\right)^2 \lesssim A_{\kappa}\left(\frac{\kappa}{\lambda}\mu\right)\lesssim 0\, .
\label{48}
\end{equation}
The experimental constraints on the SUSY parameters obtained in the MSSM remain valid in the NMSSM with small
$\lambda$ and $\kappa$. For example, non--observation of any neutral Higgs particle and chargino at the LEP II
imply that $\tan\beta\gtrsim 2.5$ and $|\mu|\gtrsim 90-100\,\mbox{GeV}$.

Decreasing $\kappa$ reduces the masses of extra scalar and pseudoscalar states so that for $\kappa\ll\lambda$
they can be the lightest particles in the Higgs boson spectrum. In the limit $\kappa\to 0$ the mass of the
lightest pseudoscalar state vanishes. In the considered limit the Lagrangian of the NMSSM is invariant under
transformations of $SU(2)\times [U(1)]^2$ global symmetry. Extra $U(1)$ global symmetry gets spontaneously
broken by the VEV of the singlet field $S$, giving rise to a massless Goldstone boson, the Peccei--Quinn (PQ)
axion. In the PQ--symmetric NMSSM astrophysical observations exclude any choice of the parameters unless
one allows $s$ to be enormously large ($>10^{9}-10^{11}\,\mbox{GeV}$). These huge vacuum expectation values of
the singlet field can be consistent with the EWSB only if $\lambda\sim 10^{-6}-10^{-9}$
\cite{Miller:2003hm}--\cite{Miller:2005qua}. Therefore here we restrict our consideration to small but non--zero 
values of $\kappa\lesssim \lambda^2$ when the PQ-symmetry is only slightly broken.

As evident from Eq.~(\ref{47}) at small values of $\kappa$ the mass--squared of the lightest Higgs scalar
tends to be negative if $|\mu|$ is large and/or the auxiliary variable $x$ differs too much from unity.
Due to the vacuum stability requirement, which implies the positivity of the mass--squared of all Higgs
particles, $x$ has to be localized near unity, i.e.
\begin{equation}
1-\left|\frac{\sqrt{2}\kappa M_Z}{\lambda^2 v}\right|<x<1+\left|\frac{\sqrt{2}\kappa M_Z}{\lambda^2 v}\right|\,.
\label{49}
\end{equation}
This leads to the hierarchical structure of the Higgs spectrum. Indeed, combining LEP limits on $\tan\beta$ and
$\mu$ one gets that $m_A^2\gtrsim 9M_Z^2\, x$. Because of this the heaviest CP--odd, CP--even and charged Higgs bosons
are almost degenerate with masses around $m_A\simeq \mu\tan\beta$ while the SM--like Higgs state has a mass of
the order of $M_Z$.

\begin{figure}
\centering
\includegraphics[width=.6\textwidth]{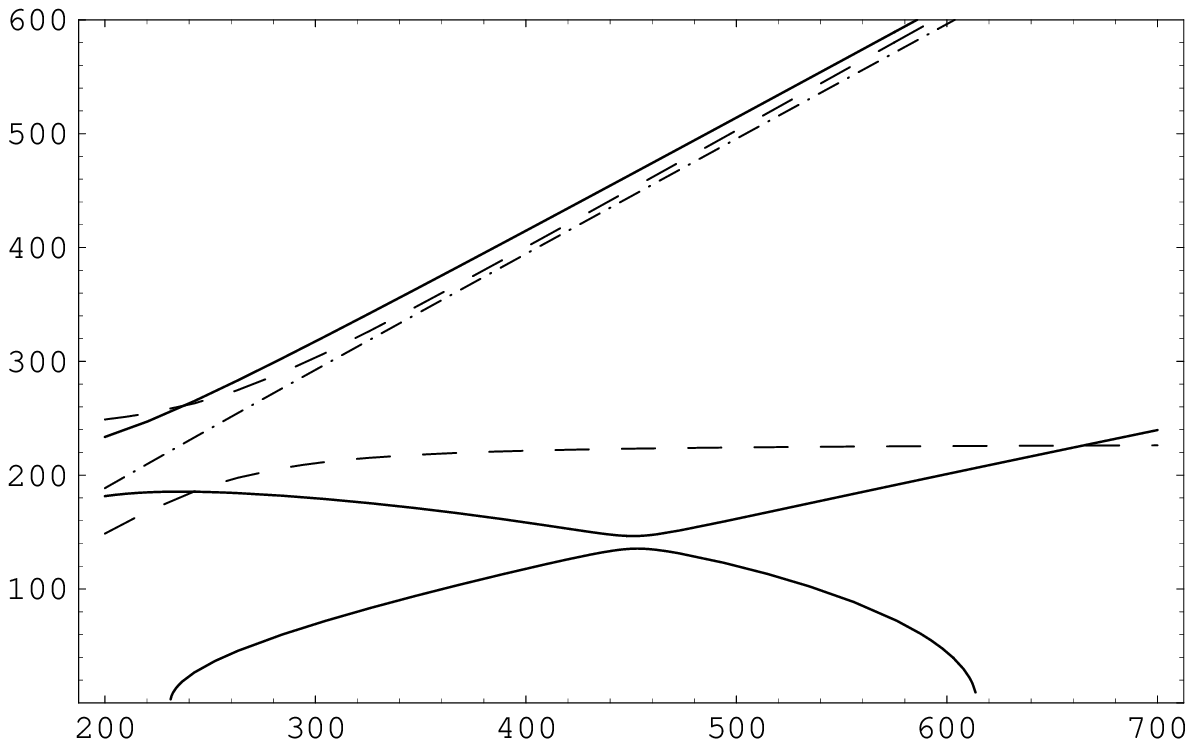}\\
{$m_A (\mbox{GeV})$}
\caption{One--loop masses of the CP--even Higgs bosons
versus $m_A$ for $\lambda=0.6$, $\kappa=0.36$, $\tan\beta=3$,
$\mu=150\,\mbox{GeV}$, $A_{\kappa}=135\,\mbox{GeV}$,
$m_Q^2=m_U^2=M_S^2$, $X_t=\sqrt{6} M_S$ and $M_S=700\,\mbox{GeV}$.
Solid, dashed and dashed--dotted lines correspond to the masses of
the CP--even, CP--odd and charged Higgs bosons respectively.}
\label{figure-2}
\end{figure}

The main features of the NMSSM Higgs spectrum discussed above are retained when the couplings
$\lambda$ and $\kappa$ increase. For the appreciable values of $\kappa$ and $\lambda$ the
slight breaking of the PQ--symmetry can be caused by
%small ratio $\kappa/\lambda$ may arise due to
the RG flow of these couplings from the GUT scale $M_X$ to $M_Z$. In the infrared region the
solutions of the NMSSM RG equations are focused near the intersection of the Hill-type effective
surface and invariant line \cite{Nevzorov:2001vj}--\cite{Nevzorov:2002ub}. As a result
at the EW scale $\kappa/\lambda$ tend to be less than unity even when $\kappa(M_X) > \lambda(M_X)$
initially. In Fig.~3 the dependence of the masses of the Higgs bosons on $m_A$ is
examined. As a representative example we fix the Yukawa couplings so that
$\lambda(M_X)=\kappa(M_X)=2 h_t(M_X)=1.6$, that corresponds to $\tan\beta\gtrsim 3$, $\lambda(M_t)=0.6$
and $\kappa(M_t)=0.36$. In order to obtain a realistic spectrum, we include the leading one--loop
corrections from the top and stop loops. From Fig.~3 it becomes clear that the requirement of
stability of the physical vacuum limits the range of variations of $m_A$ from below and above
maintaining the mass hierarchy in the Higgs spectrum. Relying on this mass hierarchy the
approximate solutions for the Higgs masses and couplings can be
obtained \cite{Miller:2003ay},\cite{Nevzorov:2004ge}. The numerical results in
Fig.~3 reveal that the masses of the heaviest CP--even, CP--odd and charged Higgs states are
approximately degenerate while the other three neutral states are considerably lighter.
The hierarchical structure of the Higgs spectrum ensures that the heaviest CP--even and CP--odd
Higgs bosons are predominantly composed of $H$ and $P$. As before the lightest Higgs scalar and
pseudoscalar are singlet dominated, making their observation quite problematic.  The second
lightest CP--even Higgs boson has a mass around $130\,\mbox{GeV}$, mimicking the lightest Higgs
scalar in the MSSM. Observing two light scalars and one pseudoscalar Higgs particles but no charged
Higgs boson at future colliders would yield an opportunity to differentiate the NMSSM with a slightly
broken PQ--symmetry from the MSSM even if the heavy Higgs states are inaccessible.

The presence of light singlet scalar and pseudoscalar permits to weaken the LEP lower bound
on the lightest Higgs boson mass. These states have reduced couplings to Z--boson that could
allow them to escape the detection at LEP. On the other hand singlet scalar can mix with
the SM-like superposition $h$ of the neutral components of Higgs doublets resulting in the
reduction of the couplings of the second lightest Higgs scalar to $Z$--boson. This relaxes
LEP constraints so that the SM-like Higgs state does not need to be considerably heavier
than $100\,\mbox{GeV}$. Therefore large contribution of loop corrections to the mass of
the SM-like Higgs boson is not required. Another possibility to overcome the little hierarchy
problem is to allow the SM--like Higgs state to decay predominantly into two light
singlet pseudoscalars $A_1$ (for recent review see \cite{Chang:2008cw}). This can be achieved
because the coupling of the SM--like Higgs boson to the $b$--quark is rather small. If this coupling
is substantially smaller than the coupling of the SM--like Higgs state to $A_1$ then the decay mode
$h\to A_1 + A_1$ dominates. The singlet pseudoscalar can sequentially decay into either
$b\bar{b}$ or $\tau\bar{\tau}$ leading to four fermion decays of the SM--like Higgs boson.
In this case, again, the corresponding Higgs eigenstate might be relatively light that
permits to avoid little hierarchy problem.

\begin{figure}
\centering
~\hspace*{-6.1cm}{$m_{h_1}$}\hspace{6.8cm}{$m_{h_1}$}\\
\includegraphics[width=0.43\textwidth]{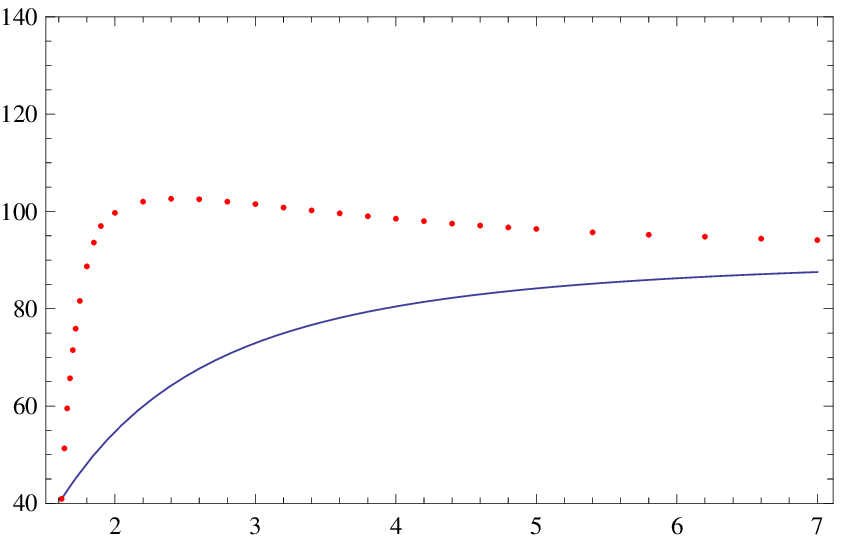}\qquad
\includegraphics[width=0.43\textwidth]{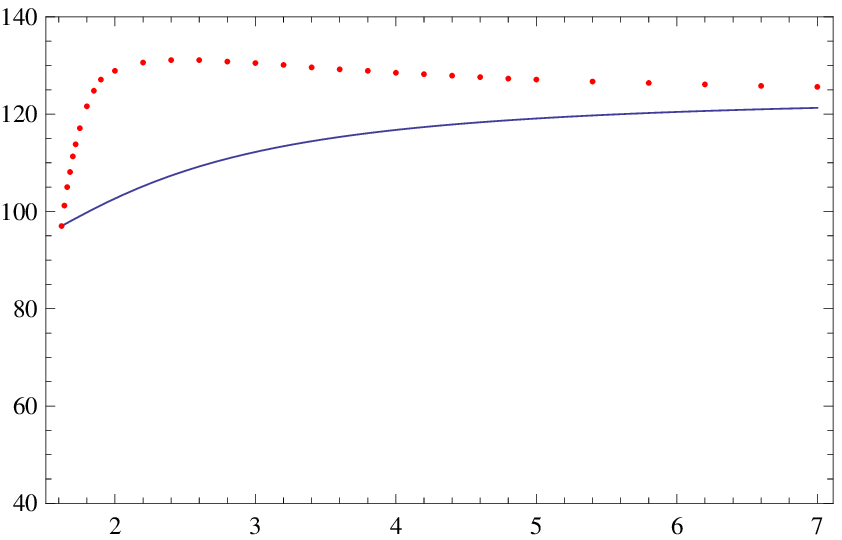}\\
~\hspace*{-0.1cm} {$\tan\beta$}\hspace{6.8cm}{$\tan\beta$}
\caption{
{\it Left:} Tree--level upper bound on the lightest Higgs boson mass
in the MSSM and NMSSM as a function of $\tan\beta$.
{\it Right:} The dependence of the two--loop upper bound on the lightest
Higgs boson mass on $\tan\beta$ for $m_t(M_t)=165\,\mbox{GeV}$,
$m_Q^2=m_U^2=M_S^2$, $X_t=\sqrt{6} M_S$ and $M_S=700\,\mbox{GeV}$.
The solid and dotted lines represent the theoretical restrictions on
$m_{h_1}$ in the MSSM and NMSSM respectively.}
\label{figure-2}
\end{figure}

However even when the couplings of the lightest CP--even Higgs state are almost the same
as in the SM it is substantially easier to overcome LEP constraint on the mass of the
SM--like Higgs boson in the NMSSM than in the MSSM. Indeed, in the NMSSM the theoretical
upper bound on $m_{h_1}^2$, which is given by $M_{22}^2$ in Eq.~(\ref{41}), contains
an extra term $\frac{\lambda^2}{2}v^2\sin^2 2\beta$ which is not present in the MSSM.
Due to this term the maximum possible value of the mass of the lightest Higgs scalar
in the NMSSM can be considerably larger as compared with the MSSM 
at moderate values of $\tan\beta$. In our analysis we
require the validity of perturbation theory up to the scale $M_X$. This sets stringent
upper limit on $\lambda(M_t)$ at low energies for each particular choice of
$\tan\beta$. Using theoretical restrictions on $\lambda(M_t)$ one can compute the
the maximum possible value of $m_{h_1}^2$ for each given value of $\tan\beta$.
Fig.~4 shows the dependence of the upper bound on the lightest Higgs boson mass as a
function of $\tan\beta$ in the MSSM and NMSSM. From Fig.~4 one can see that at the
tree--level the lightest CP--even Higgs state in the NMSSM can be considerably heavier
than in the MSSM at moderate values of $\tan\beta$. As a consequence in the leading
two--loop approximation it is substantially easier to get $m_{h_1}\gtrsim 114.4\,\mbox{GeV}$
in the NMSSM than in the MSSM for $\tan\beta=2-4$.

\subsection{Higgs spectrum in the $E_6$ inspired SUSY models with extra $U(1)'$ factor}
Another solution to the $\mu$ problem arises within superstring inspired models
based on the $E_6$ gauge group. At high energies $E_6$ can be broken
$SU(3)_C\times SU(2)_W\times U(1)_Y\times U(1)'$. An extra $U(1)'$ that appears at
low energies is a linear superposition of $U(1)_{\chi}$ and $U(1)_{\psi}$:
\begin{equation}
U(1)'=U(1)_{\chi}\cos\theta+U(1)_{\psi}\sin\theta\,,
\label{50}
\end{equation}
where two anomaly--free $U(1)_{\psi}$ and $U(1)_{\chi}$ symmetries are defined by:
$E_6\to SO(10)\times U(1)_{\psi}$\,, $SO(10)\to SU(5)\times U(1)_{\chi}$.
If $\theta\ne 0$ or $\pi$ the extra $U(1)'$ gauge symmetry forbids an elementary $\mu$
term but allows interaction $\lambda S(H_1 H_2)$ in the superpotential. After
EWSB the scalar component of the SM singlet superfield $S$ acquires a non--zero VEV
breaking $U(1)'$ and an effective $\mu$ term of the required size is automatically
generated.

The Higgs sector of the considered models includes two Higgs doublets as well
as a SM--like singlet field $S$ that carries $U(1)'$ charge. The Higgs effective
potential can be written as
\begin{equation}
\begin{array}{rcl}
V&=&V_F+V_D+V_{soft}+\Delta V\, ,\\[1mm]
V_F&=&\lambda^2|S|^2(|H_d|^2+|H_u|^2)+\lambda^2|(H_d H_u)|^2\,,\\[1mm]
V_D&=&\displaystyle\frac{g_2^2}{8}\left(H_d^\dagger \sigma_a H_d+H_u^\dagger \sigma_a
H_u\right)^2+\displaystyle\frac{{g'}^2}{8}\left(|H_d|^2-|H_u|^2\right)^2\\[2mm]
&&+\displaystyle\frac{g^{'2}_1}{2}\left(\tilde{Q}_1|H_d|^2+\tilde{Q}_2|H_u|^2+\tilde{Q}_S|S|^2\right)^2\,,\\[1mm]
V_{soft}&=&m_{S}^2|S|^2+m_1^2|H_d|^2+m_2^2|H_u|^2+
\biggl[\lambda A_{\lambda}S(H_u H_d)+h.c.\biggr]\,,
\end{array}
\label{51}
\end{equation}
where $g'_1$ is $U(1)'$ gauge coupling and $\tilde{Q}_1$, $\tilde{Q}_2$ and
$\tilde{Q}_S$ are effective $U(1)'$ charges of $H_1$, $H_2$ and $S$ respectively.
In Eq.~(\ref{51}) $V_F$ and $V_D$ are the $F$ and $D$ terms, $V_{soft}$ contains
a set of soft SUSY breaking terms while $\Delta V$ represents the contribution of
loop corrections.

At the physical vacuum the Higgs fields acquire VEVs given by Eq.~(\ref{30})
thus breaking the $SU(2)_W\times U(1)_Y\times U(1)'$ symmetry to $U(1)_{em}$. 
As a result two CP--odd and two charged Goldstone
modes in the Higgs sector are absorbed by the $Z$, $Z'$ and $W^{\pm}$ gauge bosons
so that only six physical degrees of freedom are left. They form one CP--odd, three
CP--even and two charged states. The masses of the CP--odd and charged Higgs bosons
can be written as
\begin{equation}
m_A^2=\frac{2\lambda^2 s^2 x}{\sin^2 2\beta}+O(M_Z^2)\,,\qquad\qquad m^2_{H^{\pm}}=m_A^2+O(M_Z^2)\,,
\label{52}
\end{equation}
where $x=\frac{A_{\lambda}}{\sqrt{2}\lambda s}\sin 2\beta$\,. The masses of two heaviest
CP--even states are set by $M_{Z'}$ and $m_A$, i.e.
\begin{equation}
m^2_{h_3}=m_A^2+O(M_Z^2)\,,\qquad\qquad m^2_{h_2}=M_{Z'}^2+O(M_Z^2)\,,
\label{53}
\end{equation}
where $M_{Z'}\simeq g^{'}_1\tilde{Q}_S s$. The lightest CP--even Higgs boson has a mass which
is less than
\begin{equation}
m^2_{h_1}\lesssim \frac{\lambda^2}{2}v^2\sin^22\beta+M_{Z}^2\cos^22\beta+
g^{'2}_1 v^2\biggl(\tilde{Q}_1\cos^2\beta+\tilde{Q}_2\sin^2\beta\biggr)^2+\Delta\,.
\label{54}
\end{equation}
In Eq.~(\ref{54}) $\Delta$ represents the contribution of loop corrections. Since the mass of
the $Z'$ boson in the $E_6$ inspired models has to be heavier than $800-900\,\mbox{GeV}$
at least one CP--even Higgs state, which is singlet dominated, is always heavy.
If $m_A < M_{Z'}$ then we get MSSM--type Higgs spectrum. When $m_A > M_{Z'}$
the heaviest CP--even, CP--odd and charged states are almost degenerate
with masses around $m_A$. In this case the lightest Higgs state is predominantly the
SM-like superposition $h$ of the neutral components of Higgs doublets.

Recently the detailed analysis of the Higgs sector was performed within a particular
$E_6$ inspired SUSY model with an extra $U(1)_N$ gauge symmetry that corresponds to
$\theta=\arctan\sqrt{15}$ \cite{King:2005jy}-\cite{King:2005my}.
The extra $U(1)_N$ gauge symmetry is defined such that
right--handed neutrinos do not participate in the gauge interactions. Only in this
Exceptional Supersymmetric Standard Model (E$_6$SSM) right--handed may be superheavy,
shedding light on the origin of the mass hierarchy in the lepton sector and
providing a mechanism for the generation of the baryon asymmetry in the Universe via
leptogenesis \cite{King:2008qb}. To ensure anomaly cancellation the particle content
of the E$_6$SSM is extended to include three complete fundamental $27$ representations
of $E_6$. In addition to the complete $27_i$ multiplets the low energy particle spectrum
of the E$_6$SSM is supplemented by $SU(2)_W$ doublet $H'$ and anti-doublet $\overline{H}'$
states from extra $27'$ and $\overline{27'}$ to preserve gauge coupling unification.
The unification of gauge couplings in the considered model can be achieved for any
phenomenologically acceptable value of $\alpha_3(M_Z)$ consistent with the measured low
energy central value \cite{King:2007uj}. The Higgs spectrum within the E$_6$SSM was
studied in \cite{King:2005jy}-\cite{King:2005my}, \cite{King:2006vu}--\cite{King:2006rh}.
It was argued that even at the tree level the lightest Higgs boson
mass in this model can be larger than $120\,\mbox{GeV}$. Therefore nonobservation of the
Higgs boson at LEP does not cause any trouble for the E$_6$SSM, even at tree--level.
In the leading two--loop approximation the mass of the lightest CP--even Higgs boson
in the considered model does not exceed $150-155\,\mbox{GeV}$ \cite{King:2005jy}.
The presence of light exotic particles in the E$_6$SSM spectrum lead to the nonstandard
decays of the SM--like Higgs boson which were discussed in \cite{10}.

\section*{Acknowledgements}
\vspace{-2mm} Author would like to thank E.~E.~Boos, M.~Dubinin, S.~F.~King, J.~P.~Kumar, 
D.~A.~Ross, V.~A.~Rubakov and X.~R.~Tata for fruitful discussions. Author is also
grateful to O.~Loiko, S.~Moretti, L.~B.~Okun, M.~Sher, M.~Shifman, M.~I.~Vysotsky and
P.~V.~Zinin, for valuable comments and remarks. The work of R.N. was supported by 
the U.S. Department of Energy under Contract DE-FG02-04ER41291.

\end{document}